\documentclass[aps,english,prb,floatfix,amsmath,superscriptaddress,tightenlines,twocolumn]{revtex4-2}

\usepackage{amsmath}
\usepackage{amssymb}
\usepackage{amsthm}
\usepackage{amsfonts}

\usepackage{graphicx,color,framed}
\usepackage{hyperref}
\usepackage{times}
\usepackage{enumerate}
\usepackage{lipsum}
\usepackage{slashed}
\usepackage{url}
\usepackage{comment}
\usepackage{bbm}

\usepackage[caption=false]{subfig}

\hypersetup{
    colorlinks=true, %set true if you want colored links
    linktoc=all,     %set to all if you want both sections and subsections linked
    linkcolor=blue,  %choose some color if you want links to stand out
}

\newcommand{\<}{\langle}
\renewcommand{\>}{\rangle}
\newcommand{\be}{\begin{equation} }
\newcommand{\ee}{\end{equation} }
\newcommand{\ba}{\begin{eqnarray} }
\newcommand{\ea}{\end{eqnarray} }

\newcommand{\id}{\mathbbm{1}}

\newcommand{\bpm}{\begin{pmatrix}}
\newcommand{\epm}{\end{pmatrix}}
\newcommand{\bmm}{\begin{matrix}}
\newcommand{\emm}{\end{matrix}}

\newcommand{\bea}{\begin{eqnarray}}
\newcommand{\eea}{\end{eqnarray}}

\newcommand{\beq}{\begin{equation} }
\newcommand{\eeq}{\end{equation} }
\newcommand{\beqs}{\begin{equation} \begin{split}}
\newcommand{\eeqs}{\end{split} \end{equation} }
\newcommand{\es}{\end{split}}

\newcommand{\Tr}{\text{Tr}}

\newcommand{\str}[2]{\mathbf{X}_{#1}^{#2}}
\begin{document}

%Many-body systems with spurious modular commutators
%Quantum many-body systems with spurious modular commutators
%Many-body systems with spurious values of the mdoular commutator
%Examples of many-body systems with spurious modular commutators

\title{Many-body systems with spurious modular commutators}

\author{Julian Gass}
\author{Michael Levin}
\affiliation{Kadanoff Center for Theoretical Physics, University of Chicago, Chicago, Illinois 60637,  USA}
%\date{\today}

\begin{abstract}
Recently, it was proposed that the chiral central charge of a gapped, two-dimensional quantum many-body system is proportional to a bulk ground state entanglement measure known as the modular commutator. While there is significant evidence to support this relation, we show in this paper that it is not universal. We give examples of lattice systems that have vanishing chiral central charge which nevertheless give nonzero ``spurious'' values for the modular commutator for arbitrarily large system sizes, in both one and two dimensions. Our examples are based on cluster states and utilize the fact that they can generate nonlocal modular Hamiltonians.
\end{abstract}

\maketitle

\textbf{\emph{Introduction}}---
A general property of two-dimensional (2D) gapped quantum many-body systems is that their thermal Hall conductance $\kappa_H$ is quantized at low temperatures. Specifically, for temperatures $T$ much smaller than the bulk gap, $\kappa_H$ is quantized as a rational multiple of $\pi^2 k_B^2 T/3 h$. The rational coefficient $c_-$
% -- the so-called chiral central charge --
--- known as the ``chiral central charge'' --- is an important topological invariant characterizing gapped quantum phases ~\cite{KaneFisher_1997, cappelli2002thermal, Kitaev_2006}. 

A longstanding puzzle regarding the chiral central charge is that $c_-$ is believed to be determined by the bulk ground state, but all known definitions of this quantity involve geometries with an edge at finite temperature~\cite{Kitaev_2006, KapustinSpodyneiko_2019}. This mismatch is unsatisfying from a conceptual point of view; in principle, it should be possible to compute $c_-$ from a ground state wave function on an infinite 2D plane, but it is not obvious how to do so. 

\begin{figure}
    \centering
    \def\svgwidth{0.6\columnwidth}
    %% Creator: Inkscape 1.3.2 (091e20e, 2023-11-25), www.inkscape.org
%% Accompanies image file 'ABC_new.pdf' (pdf, eps, ps)
\begingroup%
  \makeatletter%
  \providecommand\color[2][]{%
    \errmessage{(Inkscape) Color is used for the text in Inkscape, but the package 'color.sty' is not loaded}%
    \renewcommand\color[2][]{}%
  }%
  \providecommand\transparent[1]{%
    \errmessage{(Inkscape) Transparency is used (non-zero) for the text in Inkscape, but the package 'transparent.sty' is not loaded}%
    \renewcommand\transparent[1]{}%
  }%
  \providecommand\rotatebox[2]{#2}%
  \newcommand*\fsize{\dimexpr\f@size pt\relax}%
  \newcommand*\lineheight[1]{\fontsize{\fsize}{#1\fsize}\selectfont}%
  \ifx\svgwidth\undefined%
    \setlength{\unitlength}{779.52755906bp}%
    \ifx\svgscale\undefined%
      \relax%
    \else%
      \setlength{\unitlength}{\unitlength * \real{\svgscale}}%
    \fi%
  \else%
    \setlength{\unitlength}{\svgwidth}%
  \fi%
  \global\let\svgwidth\undefined%
  \global\let\svgscale\undefined%
  \makeatother%
  \begin{picture}(1,0.91636364)%
    \lineheight{1}%
    \setlength\tabcolsep{0pt}%
    \put(0,0){\includegraphics[width=\unitlength,page=1]{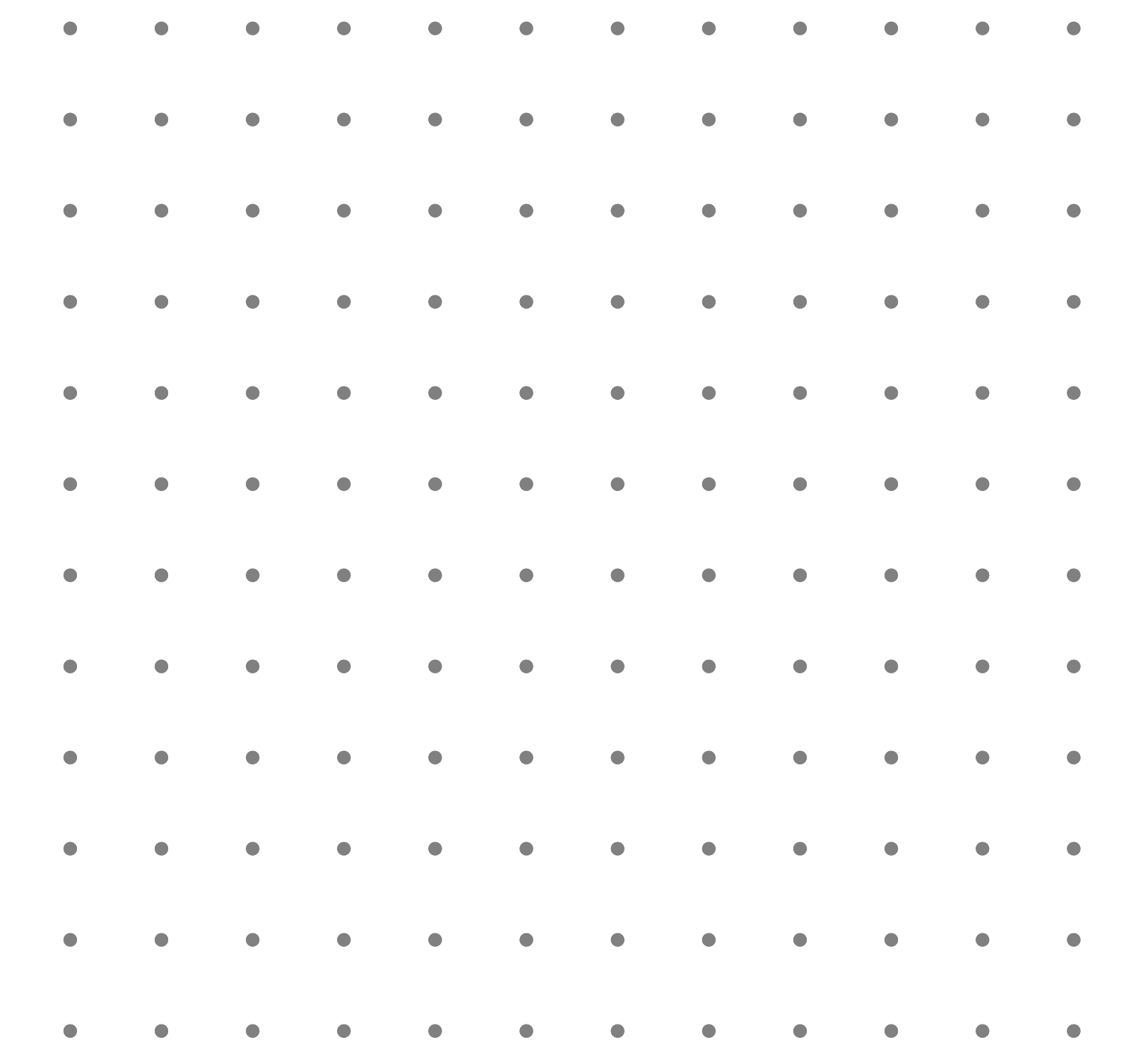}}%
    \begin{LARGE}
    	\put(0.28199002,0.52390716){\color[rgb]{0.94509804,0.04705882,0.04705882}\rotatebox{-0.00015156}{\makebox(0,0)[lt]{\lineheight{1.25}\smash{\begin{tabular}[t]{l}$A$\end{tabular}}}}}%
    	\put(0.63678415,0.52390713){\color[rgb]{0.94509804,0.04705882,0.04705882}\rotatebox{-0.00023302}{\makebox(0,0)[lt]{\lineheight{1.25}\smash{\begin{tabular}[t]{l}$C$\end{tabular}}}}}%
    	\put(0.44213726,0.22390713){\color[rgb]{0.94509804,0.04705882,0.04705882}\rotatebox{-0.00007571}{\makebox(0,0)[lt]{\lineheight{1.25}\smash{\begin{tabular}[t]{l}$B$\end{tabular}}}}}%
    \end{LARGE}
    \put(0,0){\includegraphics[width=\unitlength,page=2]{ABC_new.pdf}}%
  \end{picture}%
\endgroup%

    \caption{Geometry used to compute the modular commutator in Eq.~(\ref{eq:CCC}).}
    \label{fig:generic_ABC}
\end{figure}

Recently, Refs.~\onlinecite{Kim_2022one,Kim_2022two} proposed a new quantity --- known as the ``modular commutator’’ --- that shows promise in this direction. Let $\rho$ be a density operator defined on a 2D lattice, and let $A, B, C$ be three non-overlapping subsets of the lattice. The modular commutator, denoted $J(A,B,C)_\rho$, is defined by
\begin{align}
\label{eq:modularcommutator}
J(A,B,C)_\rho &=i\Tr\left( \rho_{ABC}[K_{AB},K_{BC}] \right),
\end{align}
where $\rho_{ABC}$ is the reduced density operator on $ABC$ and $K_R\equiv -\ln\rho_R$ is the modular Hamiltonian in region $R$. Refs.~\onlinecite{Kim_2022one,Kim_2022two} proposed that for sufficiently large $A, B, C$, arranged as in Fig.~\ref{fig:generic_ABC}, the modular commutator approaches a universal value proportional to the chiral central charge of the system:
\begin{align}
\label{eq:CCC}
J(A,B,C)_\rho=\frac{\pi}{3}c_-,
\end{align}
The evidence for Eq.~(\ref{eq:CCC}) includes (i) a physical derivation~\cite{Kim_2022one, Kim_2022two} in the case where $\rho$ satisfies the entanglement bootstrap axioms~\cite{bootstrap}, (ii) a proof in the case where $\rho$ is a gapped ground state of a quadratic fermionic Hamiltonian~\cite{Fan_2023}, and (iii) numerical results for the case of a Laughlin-like ground state~\cite{Kim_2022two} (see also Refs.~\onlinecite{Zou_mod_comm_CFT_2022, Fan_mod_comm_CFT_2022} for related CFT-based calculations). Given this evidence, one might conjecture that Eq.~(\ref{eq:CCC}) holds for \emph{all} 2D gapped ground states.

In this paper, we show that this conjecture is too strong: there exist states where the modular commutator takes on spurious values that are unrelated to $c_-$. We construct specific examples of gapped ground states that are in the same topologically trivial phase as a product state but nevertheless have a nonzero $J(A,B,C)_\rho$ even in the thermodynamic limit. These states serve as counterexamples to Eq.~(\ref{eq:CCC}), and show that the modular commutator is not a true topological invariant. 

At the same time, we also show that our examples are fine-tuned in the sense that small perturbations can cause the spurious contribution to the modular commutator to vanish in the thermodynamic limit. This situation is similar to that of the topological entanglement entropy~\cite{kitaevpreskill2006, levinwen2006}, which has also been shown to take on spurious values in certain (fine-tuned) systems~\cite{BravyiUnpublished_2008, Zou_2016, Williamson_2019, Cano_2015, Fliss_2017, Santos_2018, Stephen_2019, Kato_2020}. 
%----------------------------------------------------------------------------------------------------------------

\begin{figure}
    \centering
    \def\svgwidth{0.6\columnwidth}
    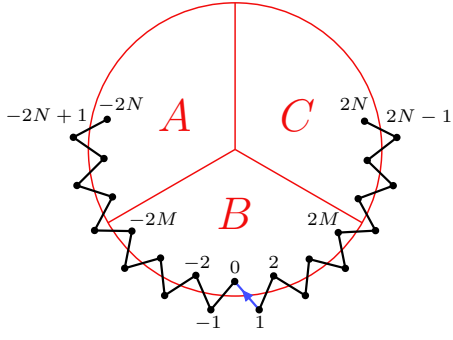
    %\includegraphics[width=0.6\columnwidth]{1dexample.pdf}
    %\caption{\JG{Depiction of our 1D system and a circuit that generates a nonzero modular commutator from a product state. Each vertex corresponds to a qubit, each black edge corresponds to a $CZ$ gate, and the oriented blue line corresponds to the two qubit gate given by (\ref{eq:Vdef}).}}
    \caption{Depiction of our 1D example violating Eq.~(\ref{eq:CCC}). Region $ABC$ consists of $2N+1$ even-numbered qubits, with region $B$ containing qubits ranging from $-2M$ to $2M$. Each vertex corresponds to a qubit while the black and blue edges represent a depth-two circuit that creates our state from a product state.}
    %(\ref{eq:circuit}).}
    \label{fig:1dsystem_and_graph}
\end{figure}

\textbf{\emph{A class of states with $J(A,B,C)_{\rho} \neq 0$}}---
%\label{sec:generalapproach}
The simplest place to look for a counterexample to Eq.~(\ref{eq:CCC}) is in a one-dimensional (1D) system, so we will begin by considering a chain of qubits arranged along the outside boundary of the region $ABC$ as in Fig.~\ref{fig:1dsystem_and_graph}. Our strategy for constructing a counterexample is to work backwards: first we identify a simple class of density operators $\rho_{ABC}$ with $J(A,B,C)_\rho \neq 0$. Then, we construct a topologically trivial many-body ground state $\rho = |\psi\>\<\psi|$ whose reduced density operator $\rho_{ABC}$ belongs to this class. 

Following this strategy, we now construct a $\rho_{ABC}$ with $J(A,B,C)_{\rho} \neq 0$. Since we clearly need $[\rho_{AB},\rho_{BC}]\neq 0$ in any such system, we start from this constraint. Recall that any density operator can be written as a linear combination of the identity operator and a collection of Pauli string operators -- tensor products of single-qubit Pauli operators $X, Y, Z$ and the identity operator $\id$. In this Pauli string representation, the simplest possibility for $\rho_{AB}, \rho_{BC}$ is that they are of the form
\begin{align}
\label{eq:pabpbcform}
    \rho_{AB}=\frac{1}{2^{N_{AB}}}(\id+\alpha P_{AB}), \quad \rho_{BC}=\frac{1}{2^{N_{BC}}}(\id+\beta P_{BC}).
\end{align}
where $P_{AB}, P_{BC}$ are two Pauli string operators, $N_R$ is the number of qubits in region $R$, and $\alpha, \beta$ are real constants with $|\alpha|, |\beta| \leq 1$. As we want $[\rho_{AB}, \rho_{BC}] \neq 0$, we need to choose $P_{AB}$ and $P_{BC}$ so that they \emph{anticommute}. The corresponding modular Hamiltonians $K_R = - \ln \rho_R$ can be computed straightforwardly:
\begin{align}
\label{eq:exampleK}
    K_{AB}
    &= c_{AB}\id - \frac{1}{2}\ln\left(\frac{1+\alpha}{1-\alpha}\right) P_{AB}, \nonumber \\
        K_{BC}&= c_{BC}\id - \frac{1}{2}\ln\left(\frac{1+\beta}{1-\beta}\right) P_{BC}.
\end{align}
where $c_{AB}, c_{BC}$ are unimportant constants. Since $P_{AB}$ and $P_{BC}$ anticommute, the commutator $[K_{AB},K_{BC}]$ is nonzero and given by
\begin{align}
\label{eq:Kcommutator}
    [K_{AB},K_{BC}]=\frac{1}{2}\ln\left(\frac{1+\alpha}{1-\alpha}\right)\ln\left(\frac{1+\beta}{1-\beta}\right)P_{AB}P_{BC}.
\end{align}

If we can now find a density operator on $ABC$ that reduces to (\ref{eq:pabpbcform}) on regions $AB$ and $BC$, and gives a nonzero expectation value for (\ref{eq:Kcommutator}), we will have a nonzero modular commutator. To this end, let us further assume that $P_{AB}$ has nontrivial support on $A$ and $P_{BC}$ has nontrivial support on $C$, so that $\Tr_AP_{AB}=\Tr_CP_{BC}=0$. With this added condition, we see immediately that the
operator
\begin{align}
\label{eq:abcForm}
    \rho_{ABC}=\frac{1}{2^{N_{ABC}}} \left[\id + (\alpha P_{AB} + \beta P_{BC}+i\gamma P_{AB}P_{BC})\right]
\end{align}
satisfies $\Tr_C \rho_{ABC} = \rho_{AB}$ and $\Tr_A \rho_{ABC} = \rho_{BC}$ as required. Furthermore, $\rho_{ABC}$ is a valid density operator as long as $\gamma$ is a real\footnote{The operator $i \gamma P_{AB} P_{BC}$ is Hermitian since $P_{AB}, P_{BC}$ anticommute.} constant satisfying $\alpha^2+\beta^2+\gamma^2\leq 1$. (This inequality ensures that $\rho_{ABC}$ is positive semi-definite since the term inside the parentheses squares to $([\alpha^2+\beta^2+\gamma^2]\cdot\id$). Finally, the presence of the $P_{AB}P_{BC}$ term ensures that the expectation value of (\ref{eq:Kcommutator}) is nonzero. Evaluating the modular commutator, we obtain:
\begin{align}
\label{eq:abcmodcom}
    J(A,B,C)_\rho
    %&=i\Tr\left(\rho_{ABC}[K_{AB},K_{BC}]\right)\nonumber\\
    &=\frac{\gamma}{2}\ln\left(\frac{1+\alpha}{1-\alpha}\right)\ln\left(\frac{1+\beta}{1-\beta}\right).
\end{align}
We conclude that any density operator of the form (\ref{eq:abcForm}) with $\alpha, \beta, \gamma \neq 0$ will have a nonzero modular commutator. 

A notable property of the above class of density operators is that at least one of the modular Hamiltonians $K_{AB}, K_{BC}$ (\ref{eq:exampleK}) is \emph{nonlocal}. To see this, recall that $P_{AB}, P_{BC}$ are assumed to have two properties: (i) they anticommute, and (ii) they have nontrivial support on $A$ and $C$ respectively. These two properties imply that at least one of $P_{AB}, P_{BC}$ is supported on a region with a diameter of order the size of region $B$. 

One might worry that the nonlocal structure of $K_{AB}, K_{BC}$ is unphysical and will prevent us from realizing (\ref{eq:abcForm}) as the reduced density operator of a gapped ground state. However, there is no such obstruction: nonlocal modular Hamiltonians of this kind are known to occur for some gapped ground states,
such as the one-dimensional cluster state~\cite{Briegel_cluster_state_2001, BravyiUnpublished_2008, Zou_2016} (see Appendix~\ref{sec:clusterstate} for a review). Building on this observation, we now construct a \emph{modified} cluster state whose reduced density operator is of the desired form (\ref{eq:abcForm}).
%----------------------------------------------------------------------------------------------------------------

\textbf{\emph{1D example}}---
%\label{sec:manybodyexample}
To define our state, we first need to set up some notation. We number our qubits from $-2N$ to $2N$ as shown in Fig.~\ref{fig:1dsystem_and_graph}. The qubits alternate between being inside and outside $ABC$, with the odd-numbered qubits lying on the outside.

The parent Hamiltonian for our state is identical to the parent Hamiltonian for the 1D cluster state with open boundary conditions except for a modification near sites $0, 1$. Specifically, our Hamiltonian is a sum of $4N+1$ local, commuting operators:
\begin{align}
\label{eq:1dex_H}
    H = -\sum_{i=-2N}^{2N}h_{i}.
\end{align}
The $h_i$ are the usual cluster state operators for all $-2N+1\leq i\leq 2N-1$: with $i \neq 0, 1$:
\begin{align}
\label{eq:hi}
    h_i = Z_{i-1}X_i Z_{i+1} \qquad (i\neq 1, 0)
\end{align}
At the two endpoints of the chain, $i = \pm 2N$, the $h_i$'s are again given by the usual cluster state operators:
\begin{align}
\label{eq:hend}
    h_{-2N} = X_{-2N}Z_{-2N+1}, \quad h_{2N} = Z_{2N-1}X_{2N}.
\end{align}
The only difference from the 1D cluster Hamiltonian is in the two terms $h_0$ and $h_1$. The $h_0$ term is defined as
\begin{align}
\label{eq:h0}
    h_0 = \frac{1}{\sqrt{3}}(Z_{-1}X_0 + Z_0Z_{1} + Z_{-1}Y_0 Z_{1}).
\end{align}
Notice that $h_0$, unlike the other $h_i$ operators, is a sum of three anticommuting Pauli string operators. As we will see below, it is this structure that leads to a ground state whose reduced density operator is of the desired form (\ref{eq:abcForm}). 
Finally, we define $h_1$ so that it commutes with the rest of the $h_i$:
\begin{align}
\label{eq:h1}
    h_{1} = Z_{-1}X_0X_{1}Z_{2}.
\end{align}

The above $h_i$'s have a number of important properties. First, $[h_i, h_j] =0$, as we mentioned earlier. Second, one can check that $h_i^2 = \id$ so each $h_i$ has eigenvalues $\pm 1$. Together these two properties imply that the eigenvalues of $H$ are integers, so in particular $H$ is gapped. These properties also imply that any ground state $|\psi\>$ of $H$ must satisfy $h_i|\psi\>=|\psi\>$ for every $i$. We can check that this ground state is \emph{unique} by considering the projector onto the ground state space 
\begin{align}
\label{eq:1dex_rho}
    \rho =\prod_{i=-2N}^{2N}\left(\frac{\id+h_i}{2}\right).
\end{align}
If $\Tr(\rho)=1$, then it follows that $H$ has a unique ground state $|\psi\>$ with $\rho = |\psi\>\<\psi|$. One can verify that this is the case by noting that every nontrivial product of the $h_i$ will itself be either a Pauli operator or a sum of Pauli operators and hence traceless. Therefore the only term in (\ref{eq:1dex_rho}) that contributes to the trace of $\rho$ is the product of the identities, which has unit trace due to the prefactor of $1/2^{4N+1}$.  

So far we have established that $H$ has a unique gapped ground state $|\psi\>$ whose density operator $\rho = |\psi\>\<\psi|$ is given by (\ref{eq:1dex_rho}). We now show that the \emph{reduced} density operator $\rho_{ABC}$ is of the desired form (\ref{eq:abcForm}). Our strategy for computing $\rho_{ABC}$ is to expand out the product in (\ref{eq:1dex_rho}) and trace out the odd-numbered qubits (since these are qubits that live outside of $ABC$). Many of the resulting terms will automatically trace to zero -- in particular, this is the case for any term that contains an odd-numbered $h_i$. Indeed, we can see from the definitions (\ref{eq:hi})-(\ref{eq:h1}), that if $i$ is odd then $h_i$ contains an (odd-numbered) $X_i$ operator which cannot be cancelled by multiplication with any other $h_j$. Therefore, the  only terms in (\ref{eq:1dex_rho}) that can survive the trace are products of even-numbered $h_i$. At the same time, the even-numbered $h_i$ with $i \neq 0$ each contain $Z$ operators acting on neighboring odd-numbered sites $i \pm 1$, so the only way that these terms can survive the trace is if they appear in products of the form $\cdots h_{2k-2} h_{2k} h_{2k+2} \cdots$. Taking into account the special form of $h_0$, one can see that these products need to be matched with one of the three terms in $h_0$ in order to cancel out all the $Z$ operators on odd-numbered sites. In this way, we find three terms that survive the trace: 
\begin{align}
\label{eq:1dstrings}
    \left(\prod_{i=1}^{N}h_{-2i}\right)Z_{-1}X_0 &= \str{-2N}{0}\nonumber\\
    Z_0Z_{1}\left(\prod_{i=1}^{N}h_{2i}\right) &= Z_0 \str{2}{2N} \nonumber\\  
    %Z_0 \str{2}{2N}X_{2}\nonumber\\
    \left(\prod_{i=1}^{N}h_{-2i}\right)Z_{-1}Y_0Z_1\left(\prod_{i=1}^{N}h_{2i}\right) &= \str{-2N}{-2} Y_0 \str{2}{2N},
\end{align}
where $\str{n}{m}$ is an abbreviated notation for a string of $X$ operators: 
\begin{align}
\label{eq:notation}
    \str{n}{m}\equiv X_n X_{n+2}\cdots X_{m-2}X_m.
\end{align}
The reduced density operator $\rho_{ABC}$ is obtained by summing the above three terms, as well as the identity operator (which also survives the trace). Including a factor of $1/\sqrt{3}$ coming from $h_0$ (\ref{eq:h0}), as well as an appropriate normalization factor, we obtain
\begin{align}
\label{eq:1dpabc}
\rho_{ABC} = \frac{1}{2^{2N+1}} \bigl[\id + \frac{1}{\sqrt{3}}\bigl(\str{-2N}{0} &+ Z_0 \str{2}{2N} \nonumber\\
&+ \str{-2N}{-2}Y_0\str{2}{2N}\bigr)\bigr].
\end{align}
We can now verify the claim we made earlier: we can see that $\rho_{ABC}$ is indeed of the form (\ref{eq:abcForm}) with $\alpha=\beta=\gamma=1/\sqrt{3}$. Therefore, the modular commutator of this state, $J(A,B,C)_\rho$, takes the nonzero value given in Eq.~(\ref{eq:abcmodcom}), independent of system size. 

To complete the discussion, we now show that $|\psi\>$ can be created from a product state by applying a depth-$2$ quantum circuit. The existence of this circuit has several implications. First, it reveals that the modular commutator is not invariant under finite-depth circuits and is therefore not a genuine topological invariant. Second, it makes it clear that $|\psi\>$ violates Eq.~(\ref{eq:CCC}) since it shows $|\psi\>$ belongs to the same phase as the product state and hence must have $c_- = 0$.\footnote{To make $c_-$ well-defined, we can imagine embedding $|\psi\>$ in a 2D qubit lattice by putting every qubit in the lattice in a product state except for a 1D chain-like subset of $4N+1$ qubits which we put in the state $|\psi\>$.} 

A pictorial representation of the depth-$2$ circuit that creates $|\psi\>$ from a product state is shown in Fig.~\ref{fig:1dsystem_and_graph}. Each edge represents a two-qubit gate: the black edges denote $CZ$ gates while the single oriented blue edge denotes a special two qubit gate $V_{1,0}$ acting on sites $1$ and $0$. Here, $V_{i,j} \equiv \text{CNOT}_{i,j} \mathbf{R}_j$ where $\text{CNOT}_{i,j}$ is the controlled NOT operator with $i$ as the control qubit, and $\mathbf{R}_j$ is any single qubit rotation acting on qubit $j$ such that $\mathbf{R}_j X_j \mathbf{R}_j^\dagger = \frac{1}{\sqrt{3}}(X_j + Y_j + Z_j)$. All the gates commute with one another except for the $V$ gate, which does not commute with neighboring $CZ$ gates and is applied first. This circuit acts on the product state $|\psi_+\>$ defined by $X_i |\psi_+\> = |\psi_+\>$ for all $i$. A derivation of this circuit is given in Appendix~\ref{app:1D_circuit}. 

Using the above circuit language, we can easily generalize our example by replacing $V_{1,0}$ with another two-qubit gate. Our expectation is that a generic (time-reversal breaking) choice of this gate will still give a nonzero $J(A,B,C)_\rho$ in the thermodynamic limit. In other words, the particular way that we modified the cluster state near sites $0, 1$ is likely not important other than the fact that it breaks time-reversal symmetry and is easy to analyze.
%----------------------------------------------------------------------------------------------------------------

\begin{figure}
    \centering
    \def\svgwidth{0.7\columnwidth}
    %% Creator: Inkscape 1.3.2 (091e20e, 2023-11-25), www.inkscape.org
%% Accompanies image file '2dhexagonexample1_new.pdf' (pdf, eps, ps)
\begingroup%
  \makeatletter%
  \providecommand\color[2][]{%
    \errmessage{(Inkscape) Color is used for the text in Inkscape, but the package 'color.sty' is not loaded}%
    \renewcommand\color[2][]{}%
  }%
  \providecommand\transparent[1]{%
    \errmessage{(Inkscape) Transparency is used (non-zero) for the text in Inkscape, but the package 'transparent.sty' is not loaded}%
    \renewcommand\transparent[1]{}%
  }%
  \providecommand\rotatebox[2]{#2}%
  \newcommand*\fsize{\dimexpr\f@size pt\relax}%
  \newcommand*\lineheight[1]{\fontsize{\fsize}{#1\fsize}\selectfont}%
  \ifx\svgwidth\undefined%
    \setlength{\unitlength}{1509.68117541bp}%
    \ifx\svgscale\undefined%
      \relax%
    \else%
      \setlength{\unitlength}{\unitlength * \real{\svgscale}}%
    \fi%
  \else%
    \setlength{\unitlength}{\svgwidth}%
  \fi%
  \global\let\svgwidth\undefined%
  \global\let\svgscale\undefined%
  \makeatother%
  \begin{picture}(1,0.69204561)%
    \lineheight{1}%
    \setlength\tabcolsep{0pt}%
    \put(0,0){\includegraphics[width=\unitlength,page=1]{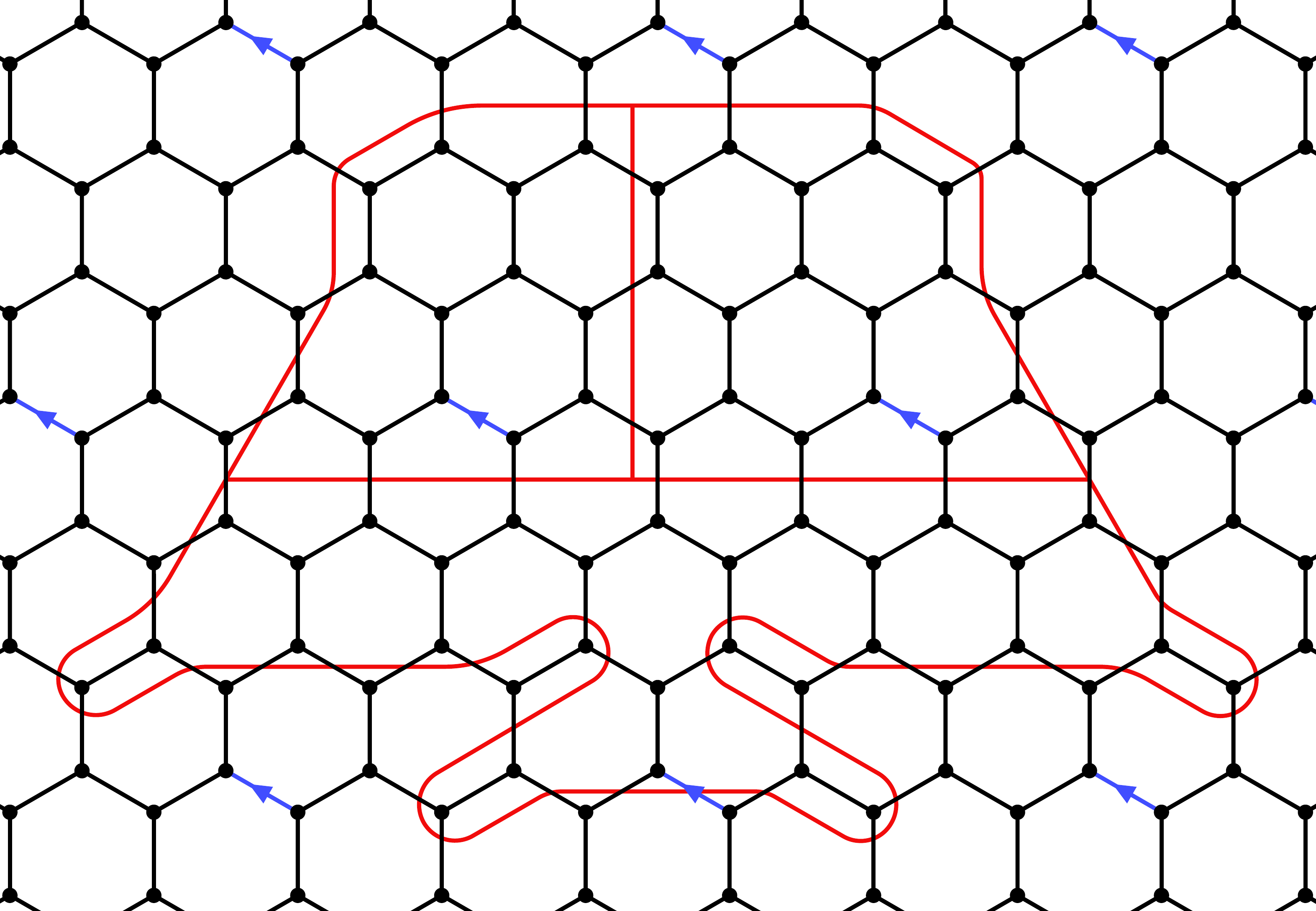}}%
    \begin{large}
    	\put(0.47107584,0.21203033){\color[rgb]{0.94509804,0.04705882,0.04705882}\makebox(0,0)[lt]{\lineheight{1.25}\smash{\begin{tabular}[t]{l}$B$\end{tabular}}}}%
    	\put(0.5848408,0.4073508){\color[rgb]{0.94509804,0.04705882,0.04705882}\makebox(0,0)[lt]{\lineheight{1.25}\smash{\begin{tabular}[t]{l}$C$\end{tabular}}}}%
    	\put(0.36364726,0.4073508){\color[rgb]{0.94509804,0.04705882,0.04705882}\makebox(0,0)[lt]{\lineheight{1.25}\smash{\begin{tabular}[t]{l}$A$\end{tabular}}}}%
    \end{large}
  \end{picture}%
\endgroup%

    \caption{Translationally invariant 2D system that violates Eq.~(\ref{eq:CCC}) for the $ABC$ configuration shown. The trapezoidal region can be made arbitrarily large, while a protrusion at the bottom is drawn so that the outer boundary intersects a single blue edge.}
    \label{fig:2dsystem}
\end{figure}
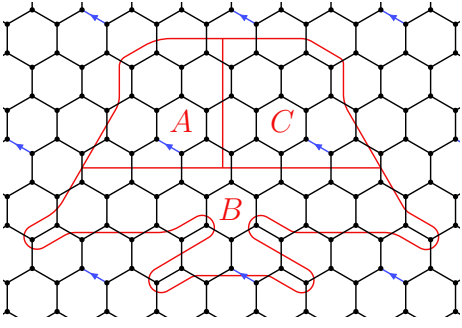

\textbf{\emph{2D example}}---
%\label{sec:2d}
We now present a second counterexample to Eq.~(\ref{eq:CCC}). This example has the advantage of being defined in a two-dimensional, translationally invariant setting, namely a honeycomb lattice with one qubit on each lattice site. As in the 1D case, our counterexample state $|\psi\>$ can be obtained by applying a finite-depth quantum circuit to the product state $|\psi_+\>$ defined by $X_i |\psi_+\> = |\psi_+\>$ for all $i$. This circuit is shown in Fig.~\ref{fig:2dsystem}. Again, the black edges represent (commuting) CZ gates while the blue edges are the two qubit $V$ gates defined above, which are applied first. Using this circuit, one can easily write down a gapped, commuting, translationally invariant parent Hamiltonian for $|\psi\>$. This parent Hamiltonian $H$ is identical to that of the cluster state on the 2D honeycomb lattice, except for a modification near the blue edges, i.e.~$H = -\sum_i h_i$, where $h_i = X_i \prod_{\<ij\>} Z_j$ for all sites $i$ that are not adjacent to a blue edge. For each pair of sites $i_0, i_1$ that are connected by a blue edge, we have $h_{i_0} = \frac{1}{\sqrt{3}}[X_{i_0}\prod_{\<ii_0\>, i\neq i_1}Z_i + Y_{i_0}\prod_{\<ii_0\>} Z_i + Z_{i_0} Z_{i_1}]$ and $h_{i_1} = X_{i_0} X_{i_1} \prod_{\<i i_0\>, i \neq i_1} Z_i \prod_{\<i i_1\>, i \neq i_0} Z_i$. The existence of this circuit also implies that $|\psi\>$ has $c_- = 0$. Nevertheless, we will now show that there exist arbitrarily large regions $A,B,C$ such that $J(A,B,C)_{|\psi\>}$ is nonzero and size independent.

\begin{figure}
\begin{footnotesize}
\subfloat[\label{fig:2dreduction1}]{\def\svgwidth{4.25cm}%% Creator: Inkscape 1.3.2 (091e20e, 2023-11-25), www.inkscape.org
%% Accompanies image file '2dhexagonexample2_new.pdf' (pdf, eps, ps)
\begingroup%
  \makeatletter%
  \providecommand\color[2][]{%
    \errmessage{(Inkscape) Color is used for the text in Inkscape, but the package 'color.sty' is not loaded}%
    \renewcommand\color[2][]{}%
  }%
  \providecommand\transparent[1]{%
    \errmessage{(Inkscape) Transparency is used (non-zero) for the text in Inkscape, but the package 'transparent.sty' is not loaded}%
    \renewcommand\transparent[1]{}%
  }%
  \providecommand\rotatebox[2]{#2}%
  \newcommand*\fsize{\dimexpr\f@size pt\relax}%
  \newcommand*\lineheight[1]{\fontsize{\fsize}{#1\fsize}\selectfont}%
  \ifx\svgwidth\undefined%
    \setlength{\unitlength}{1509.68117541bp}%
    \ifx\svgscale\undefined%
      \relax%
    \else%
      \setlength{\unitlength}{\unitlength * \real{\svgscale}}%
    \fi%
  \else%
    \setlength{\unitlength}{\svgwidth}%
  \fi%
  \global\let\svgwidth\undefined%
  \global\let\svgscale\undefined%
  \makeatother%
  \begin{picture}(1,0.69204561)%
    \lineheight{1}%
    \setlength\tabcolsep{0pt}%
    \put(0,0){\includegraphics[width=\unitlength,page=1]{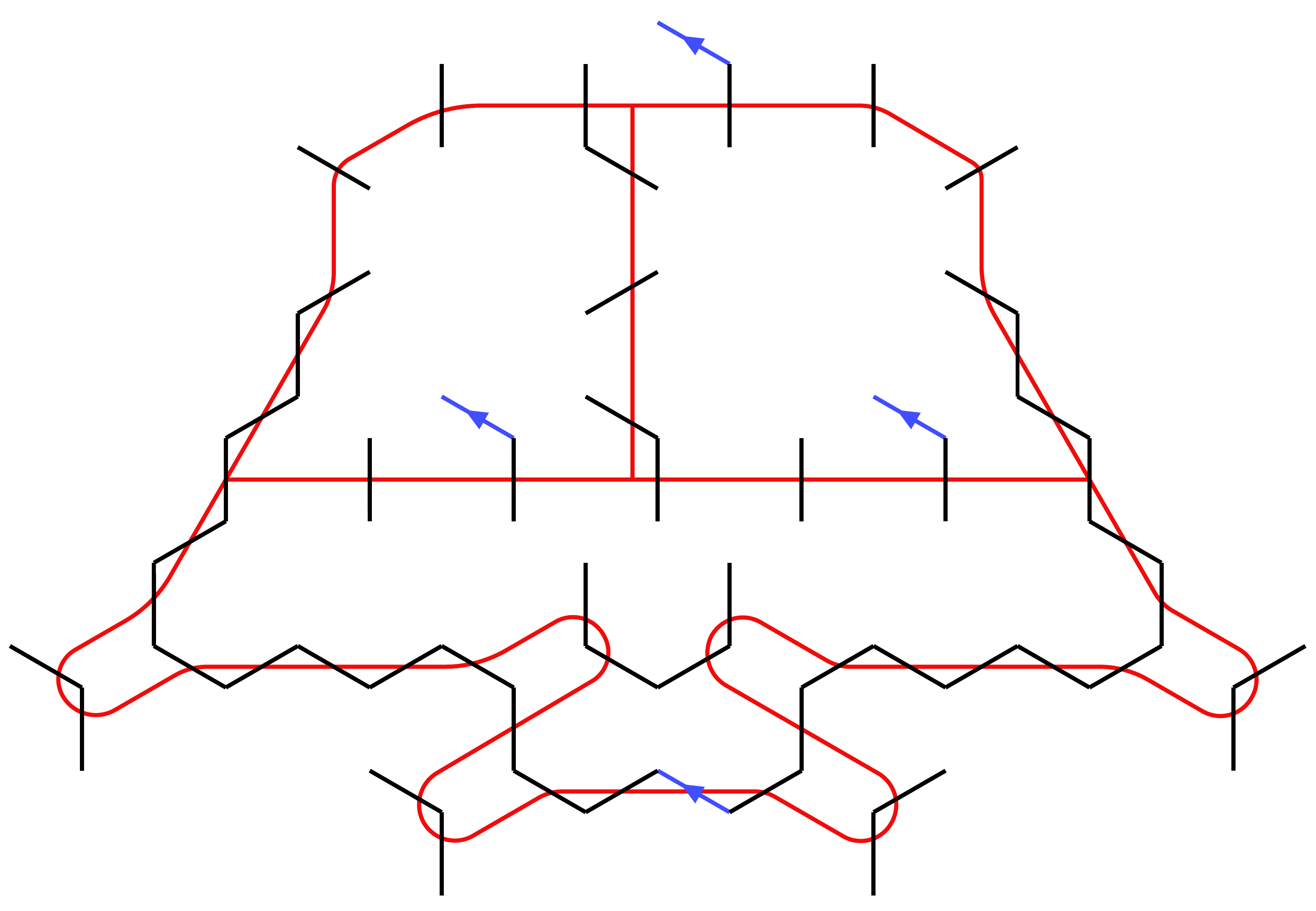}}%
    \put(0.32864726,0.4373508){\color[rgb]{0.94509804,0.04705882,0.04705882}\makebox(0,0)[lt]{\lineheight{1.25}\smash{\begin{tabular}[t]{l}$A$\end{tabular}}}}%
    \put(0.58484079,0.4373508){\color[rgb]{0.94509804,0.04705882,0.04705882}\makebox(0,0)[lt]{\lineheight{1.25}\smash{\begin{tabular}[t]{l}$C$\end{tabular}}}}%
    \put(0.47607583,0.22203033){\color[rgb]{0.94509804,0.04705882,0.04705882}\makebox(0,0)[lt]{\lineheight{1.25}\smash{\begin{tabular}[t]{l}$B$\end{tabular}}}}%
    \put(0,0){\includegraphics[width=\unitlength,page=2]{2dhexagonexample2_new.pdf}}%
  \end{picture}%
\endgroup%
}
\hfill
\subfloat[\label{fig:2dreduction2}]{\def\svgwidth{4.25cm}%% Creator: Inkscape 1.3.2 (091e20e, 2023-11-25), www.inkscape.org
%% Accompanies image file '2dhexagonexample3_new.pdf' (pdf, eps, ps)
\begingroup%
  \makeatletter%
  \providecommand\color[2][]{%
    \errmessage{(Inkscape) Color is used for the text in Inkscape, but the package 'color.sty' is not loaded}%
    \renewcommand\color[2][]{}%
  }%
  \providecommand\transparent[1]{%
    \errmessage{(Inkscape) Transparency is used (non-zero) for the text in Inkscape, but the package 'transparent.sty' is not loaded}%
    \renewcommand\transparent[1]{}%
  }%
  \providecommand\rotatebox[2]{#2}%
  \newcommand*\fsize{\dimexpr\f@size pt\relax}%
  \newcommand*\lineheight[1]{\fontsize{\fsize}{#1\fsize}\selectfont}%
  \ifx\svgwidth\undefined%
    \setlength{\unitlength}{1509.68117541bp}%
    \ifx\svgscale\undefined%
      \relax%
    \else%
      \setlength{\unitlength}{\unitlength * \real{\svgscale}}%
    \fi%
  \else%
    \setlength{\unitlength}{\svgwidth}%
  \fi%
  \global\let\svgwidth\undefined%
  \global\let\svgscale\undefined%
  \makeatother%
  \begin{picture}(1,0.69204561)%
    \lineheight{1}%
    \setlength\tabcolsep{0pt}%
    \put(0,0){\includegraphics[width=\unitlength,page=1]{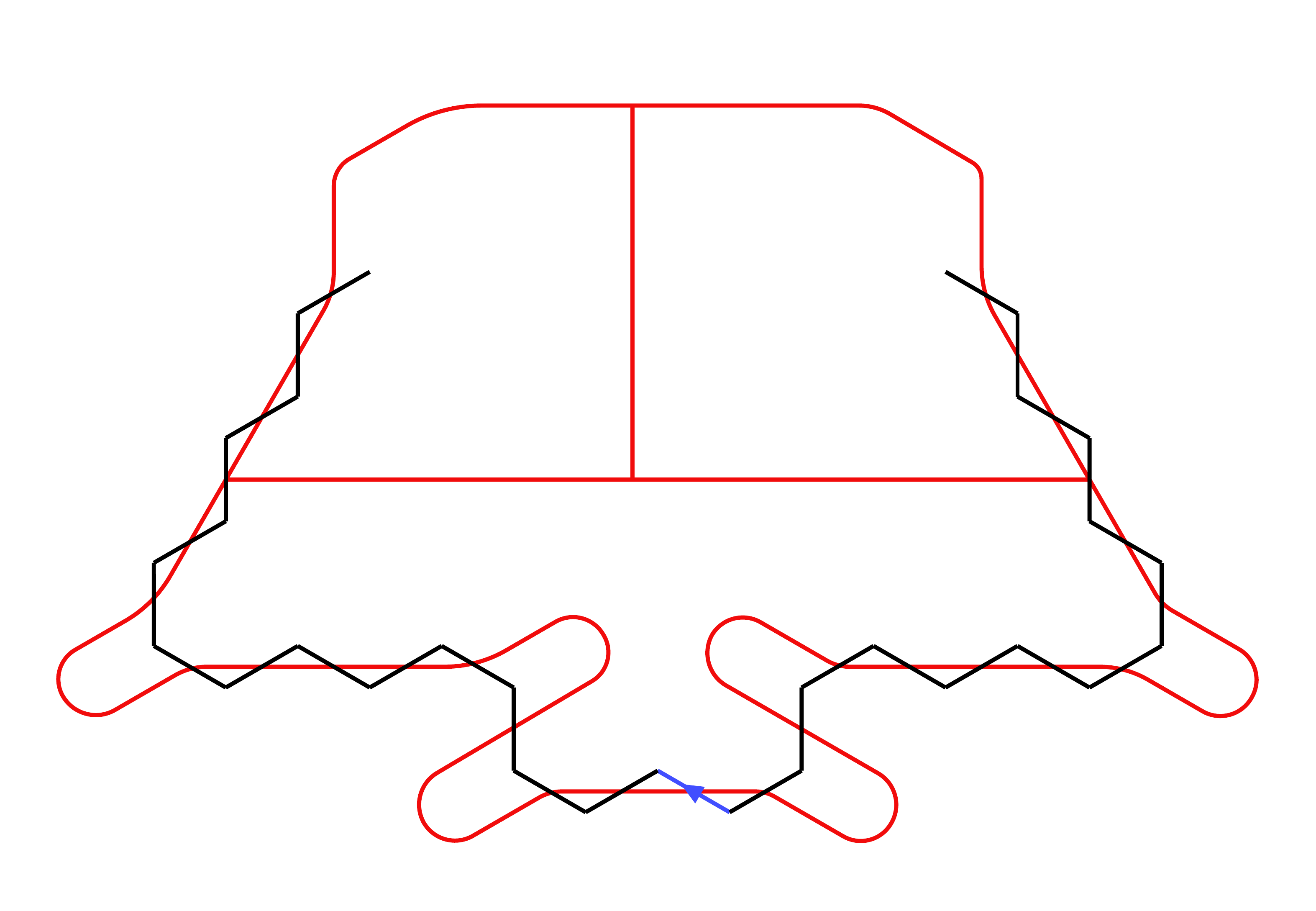}}%
    \put(0.32864726,0.4373508){\color[rgb]{0.94509804,0.04705882,0.04705882}\makebox(0,0)[lt]{\lineheight{1.25}\smash{\begin{tabular}[t]{l}$A$\end{tabular}}}}%
    \put(0.58484079,0.4373508){\color[rgb]{0.94509804,0.04705882,0.04705882}\makebox(0,0)[lt]{\lineheight{1.25}\smash{\begin{tabular}[t]{l}$C$\end{tabular}}}}%
    \put(0.47607583,0.22203033){\color[rgb]{0.94509804,0.04705882,0.04705882}\makebox(0,0)[lt]{\lineheight{1.25}\smash{\begin{tabular}[t]{l}$B$\end{tabular}}}}%
    \put(0,0){\includegraphics[width=\unitlength,page=2]{2dhexagonexample3_new.pdf}}%
  \end{picture}%
\endgroup%
}
%\subfloat[\label{fig:2dreduction1}]{\includegraphics[width=4.25cm]{2dhexagonexample2.pdf}}
%\subfloat[\label{fig:2dreduction2}]{\includegraphics[width=4.25cm]{2dhexagonexample3.pdf}}
     \caption{Reduction of $|\psi\>$ to equivalent states that consist of (a) chains that have support crossing a boundary of $ABC$ and (b) chains that have support on all regions $A, B, C, D$.}
     %\JG{showing reduction...}(a) Gates that generate entanglement across the boundaries of $ABC$. (b) The only chain that has support on all regions $A, B, C, D$.}
     \label{fig:2dreduction}
\end{footnotesize}
\end{figure}

Our argument follows from general facts of modular commutators which are given in Ref.~\onlinecite{Kim_2022two}. The first fact is that $J(A,B,C)_{|\psi\>}$ is unchanged after acting on $|\psi\>$ with any unitary operator that is supported entirely in regions $A$, $B$, $C$, or $D\equiv (ABC)^c$. This means that we can remove any entangling gate that does not intersect a boundary of $A$, $B$, or $C$ while computing the modular commutator, effectively converting our two-dimensional problem into a one-dimensional problem.\footnote{More precisely, since the blue gates are applied before the black gates, we need to remove them in the reverse order: first, we remove the \emph{black} gates that do not intersect a boundary of $A$, $B$, $C$. Then we remove the blue gates that do not intersect one of these boundaries and do not touch a residual black gate.} This reduction is depicted in Fig.~\ref{fig:2dreduction1}, where we see that the $ABC$ geometry of Fig.~\ref{fig:2dsystem} leaves us with an equivalent state $|\psi'\>$ which is the tensor product of several decoupled 1D chains. We now use the fact that the modular commutator is additive under tensor products and zero for states that have trivial support on either $A$, $B$, $C$, or $D$. Every individual chain in Fig.~\ref{fig:2dreduction1} therefore contributes separately to $J(A,B,C)_{|\psi'\>}$, and the only nonzero contributions come from the chains that are supported on all four regions. There is only one such chain, so we see that $J(A,B,C)_{|\psi'\>}=J(A,B,C)_{|\psi''\>}$, where $|\psi''\>$ is depicted in Fig.~\ref{fig:2dreduction2}. 

Comparing the circuit in Fig.~\ref{fig:2dreduction2} to the 1D circuit in Fig.~\ref{fig:1dsystem_and_graph}, we see that $|\psi''\>$ is identical to the 1D example studied earlier. Hence, the modular commutator $J(A,B,C)_{|\psi\>}$ for this choice of $A, B, C$ is again nonzero and given by (\ref{eq:abcmodcom}) for $\alpha=\beta=\gamma=1/\sqrt{3}$. Note that we can get this same value for arbitrarily large regions $A$, $B$ and $C$, since extending the side lengths of the trapezoid in Fig.~\ref{fig:2dsystem} will have no effect on the value of the modular commutator as long as the outer chain in Fig.~\ref{fig:2dreduction2} extends from region $A$ to $C$ and intersects a single blue edge in region $B$.

We note that in both the 1D and 2D examples, the modular commutator depends sensitively on the precise choice of the boundaries of $A, B, C$. For example, in the 1D system, moving any single qubit in $ABC$ to the outside of $ABC$ will result in a vanishing modular commutator. Our 2D example displays a similar sensitivity. 

%----------------------------------------------------------------------------------------------------------------

\textbf{\emph{Instability}}---
%\label{sec:stability}
An important property of our examples is that they are \emph{fine-tuned} in the sense that arbitrarily small perturbations of the parent Hamiltonian can lead to a modular commutator that vanishes in the thermodynamic limit. Therefore, while our examples violate Eq.~(\ref{eq:CCC}), this violation disappears under small perturbations. We now demonstrate this instability for our 1D example. 

We focus on a particular perturbation that is analytically tractable. Our perturbation acts on qubits in the interval $[-2M,2M]$ (see Fig.~\ref{fig:1dsystem_and_graph}) and is defined in terms of a unitary transformation $U(\theta)$ that depends on a real parameter $\theta$:
\begin{align}
    U(\theta) = \prod_{j=1}^{M}R_{-2j}(\theta) \prod_{k=1}^{M}R_{2k-1}(\theta),
    \label{eq:Utheta}
\end{align}
where $R_i(\theta)$ are the two-qubit unitaries
\begin{align}
    R_i(\theta) \equiv \text{exp}\left(-i\frac{\theta}{2}Z_iZ_{i+1}\right).
\end{align}
We imagine conjugating
%perturbing 
the parent Hamiltonian $H$ by the above unitary transformation:
\begin{align}
H \rightarrow \widetilde{H} = U(\theta) H U^\dagger(\theta)
\end{align}
Notice that $\widetilde{H}$ is a local Hamiltonian since $U(\theta)$ is a product of commuting local unitary gates. Also $\widetilde{H}$ reduces to $H$ as $\theta$ approaches $0$, so this is a valid perturbation of our parent Hamiltonian in the limit of small $\theta$.

Clearly the perturbed ground state is $|\tilde{\psi}\> = U(\theta)|\psi\>$ so the perturbed density operator is $\tilde{\rho} = U(\theta)\rho U^\dagger(\theta)$. Equivalently, using (\ref{eq:1dex_rho}), we can write $\tilde{\rho}$ as
\begin{align}
    \tilde{\rho} = \prod_{i=-2N}^{2N}\left(\frac{\id+\tilde{h}_i}{2}\right).
    \label{eq:1dex_rho_pert}
\end{align}
where $\tilde{h}_i \equiv U h_i U^\dagger$.

The modular commutator for $\tilde{\rho}$ can now be computed in the same way as in the unperturbed system. A straightforward calculation (Appendix~\ref{app:stability}) gives:
\begin{align}
\label{eq:modcomperturbed}
    J(A,B,C)_{\tilde{\rho}} 
    &=\frac{\cos^{4M}\theta}{6\sqrt{3}\delta^2}\ln^2\left(\frac{1+\delta}{1-\delta}\right)
\end{align}
where
\begin{align*}
    \delta= \sqrt{1-\frac{2}{3}\cos^{2M}\theta}
\end{align*}
From this expression we see that the modular commutator $J(A,B,C)_{\tilde{\rho}}$ is exponentially suppressed in the thermodynamic limit $M,N \rightarrow \infty$ for arbitrarily small but nonzero $\theta$. This is the instability we claimed earlier. 

%----------------------------------------------------------------------------------------------------------------

\textbf{\emph{Discussion}}---
%\label{sec:discussion}
It is notable that cluster states not only form the basis of our examples, but are also known to give spurious values for the topological entanglement entropy~\cite{BravyiUnpublished_2008, Zou_2016, Williamson_2019}. One explanation for why cluster states are associated with both types of spurious quantities is that these states have nonlocal modular Hamiltonians. Indeed, as we emphasized above, the spurious modular commutator in our examples is deeply connected to the nonlocality of their modular Hamiltonians $K_{AB}, K_{BC}$. It would be interesting to understand whether all counterexamples to Eq.~(\ref{eq:CCC}) share this nonlocal structure. 

Another important question is whether all counterexamples are as fragile as the ones we have identified here, i.e.~whether the spurious modular commutator is always unstable to small perturbations. If so, then there may be a sense in which Eq.~(\ref{eq:CCC}) holds for ``generic’’ ground states, even though it does not hold universally.
%----------------------------------------------------------------------------------------------------------------
%----------------------------------------------------------------------------------------------------------------
%----------------------------------------------------------------------------------------------------------------

\acknowledgments

J.G. and M.L. acknowledge the support of the Kadanoff Center for Theoretical Physics at the University of Chicago. This work was supported in part by a Simons Investigator grant (M.L.) and by the Simons Collaboration on Ultra-Quantum Matter, which is a grant from 
the Simons Foundation (651440, M.L.).

%----------------------------------------------------------------------------------------------------------------
%----------------------------------------------------------------------------------------------------------------
%----------------------------------------------------------------------------------------------------------------

%%%%%%%%%%%%%%%%%%%%%%%%%%%%%%%%
%%%%%%%%%%%%%%%%%%%%%%%%%%%%%%%%

%\bibliographystyle{plain}
\bibliography{modular_commutator}%{}

\begin{thebibliography}{24}%
\makeatletter
\providecommand \@ifxundefined [1]{%
 \@ifx{#1\undefined}
}%
\providecommand \@ifnum [1]{%
 \ifnum #1\expandafter \@firstoftwo
 \else \expandafter \@secondoftwo
 \fi
}%
\providecommand \@ifx [1]{%
 \ifx #1\expandafter \@firstoftwo
 \else \expandafter \@secondoftwo
 \fi
}%
\providecommand \natexlab [1]{#1}%
\providecommand \enquote  [1]{``#1''}%
\providecommand \bibnamefont  [1]{#1}%
\providecommand \bibfnamefont [1]{#1}%
\providecommand \citenamefont [1]{#1}%
\providecommand \href@noop [0]{\@secondoftwo}%
\providecommand \href [0]{\begingroup \@sanitize@url \@href}%
\providecommand \@href[1]{\@@startlink{#1}\@@href}%
\providecommand \@@href[1]{\endgroup#1\@@endlink}%
\providecommand \@sanitize@url [0]{\catcode `\\12\catcode `\$12\catcode
  `\&12\catcode `\#12\catcode `\^12\catcode `\_12\catcode `\%12\relax}%
\providecommand \@@startlink[1]{}%
\providecommand \@@endlink[0]{}%
\providecommand \url  [0]{\begingroup\@sanitize@url \@url }%
\providecommand \@url [1]{\endgroup\@href {#1}{\urlprefix }}%
\providecommand \urlprefix  [0]{URL }%
\providecommand \Eprint [0]{\href }%
\providecommand \doibase [0]{https://doi.org/}%
\providecommand \selectlanguage [0]{\@gobble}%
\providecommand \bibinfo  [0]{\@secondoftwo}%
\providecommand \bibfield  [0]{\@secondoftwo}%
\providecommand \translation [1]{[#1]}%
\providecommand \BibitemOpen [0]{}%
\providecommand \bibitemStop [0]{}%
\providecommand \bibitemNoStop [0]{.\EOS\space}%
\providecommand \EOS [0]{\spacefactor3000\relax}%
\providecommand \BibitemShut  [1]{\csname bibitem#1\endcsname}%
\let\auto@bib@innerbib\@empty
%</preamble>
\bibitem [{\citenamefont {Kane}\ and\ \citenamefont
  {Fisher}(1997)}]{KaneFisher_1997}%
  \BibitemOpen
  \bibfield  {author} {\bibinfo {author} {\bibfnamefont {C.~L.}\ \bibnamefont
  {Kane}}\ and\ \bibinfo {author} {\bibfnamefont {M.~P.~A.}\ \bibnamefont
  {Fisher}},\ }\bibfield  {title} {\bibinfo {title} {Quantized thermal
  transport in the fractional quantum hall effect},\ }\href
  {https://doi.org/10.1103/PhysRevB.55.15832} {\bibfield  {journal} {\bibinfo
  {journal} {Phys. Rev. B}\ }\textbf {\bibinfo {volume} {55}},\ \bibinfo
  {pages} {15832} (\bibinfo {year} {1997})}\BibitemShut {NoStop}%
\bibitem [{\citenamefont {Cappelli}\ \emph {et~al.}(2002)\citenamefont
  {Cappelli}, \citenamefont {Huerta},\ and\ \citenamefont
  {Zemba}}]{cappelli2002thermal}%
  \BibitemOpen
  \bibfield  {author} {\bibinfo {author} {\bibfnamefont {A.}~\bibnamefont
  {Cappelli}}, \bibinfo {author} {\bibfnamefont {M.}~\bibnamefont {Huerta}},\
  and\ \bibinfo {author} {\bibfnamefont {G.~R.}\ \bibnamefont {Zemba}},\
  }\bibfield  {title} {\bibinfo {title} {Thermal transport in chiral conformal
  theories and hierarchical quantum hall states},\ }\href@noop {} {\bibfield
  {journal} {\bibinfo  {journal} {Nuclear Physics B}\ }\textbf {\bibinfo
  {volume} {636}},\ \bibinfo {pages} {568} (\bibinfo {year}
  {2002})}\BibitemShut {NoStop}%
\bibitem [{\citenamefont {Kitaev}(2006)}]{Kitaev_2006}%
  \BibitemOpen
  \bibfield  {author} {\bibinfo {author} {\bibfnamefont {A.}~\bibnamefont
  {Kitaev}},\ }\bibfield  {title} {\bibinfo {title} {Anyons in an exactly
  solved model and beyond},\ }\href {https://doi.org/10.1016/j.aop.2005.10.005}
  {\bibfield  {journal} {\bibinfo  {journal} {Annals of Physics}\ }\textbf
  {\bibinfo {volume} {321}},\ \bibinfo {pages} {2–111} (\bibinfo {year}
  {2006})}\BibitemShut {NoStop}%
\bibitem [{\citenamefont {Kapustin}\ and\ \citenamefont
  {Spodyneiko}(2020)}]{KapustinSpodyneiko_2019}%
  \BibitemOpen
  \bibfield  {author} {\bibinfo {author} {\bibfnamefont {A.}~\bibnamefont
  {Kapustin}}\ and\ \bibinfo {author} {\bibfnamefont {L.}~\bibnamefont
  {Spodyneiko}},\ }\bibfield  {title} {\bibinfo {title} {Thermal hall
  conductance and a relative topological invariant of gapped two-dimensional
  systems},\ }\href {https://doi.org/10.1103/PhysRevB.101.045137} {\bibfield
  {journal} {\bibinfo  {journal} {Phys. Rev. B}\ }\textbf {\bibinfo {volume}
  {101}},\ \bibinfo {pages} {045137} (\bibinfo {year} {2020})}\BibitemShut
  {NoStop}%
\bibitem [{\citenamefont {Kim}\ \emph {et~al.}(2022{\natexlab{a}})\citenamefont
  {Kim}, \citenamefont {Shi}, \citenamefont {Kato},\ and\ \citenamefont
  {Albert}}]{Kim_2022one}%
  \BibitemOpen
  \bibfield  {author} {\bibinfo {author} {\bibfnamefont {I.~H.}\ \bibnamefont
  {Kim}}, \bibinfo {author} {\bibfnamefont {B.}~\bibnamefont {Shi}}, \bibinfo
  {author} {\bibfnamefont {K.}~\bibnamefont {Kato}},\ and\ \bibinfo {author}
  {\bibfnamefont {V.~V.}\ \bibnamefont {Albert}},\ }\bibfield  {title}
  {\bibinfo {title} {Chiral central charge from a single bulk wave function},\
  }\href {https://doi.org/10.1103/PhysRevLett.128.176402} {\bibfield  {journal}
  {\bibinfo  {journal} {Phys. Rev. Lett.}\ }\textbf {\bibinfo {volume} {128}},\
  \bibinfo {pages} {176402} (\bibinfo {year} {2022}{\natexlab{a}})}\BibitemShut
  {NoStop}%
\bibitem [{\citenamefont {Kim}\ \emph {et~al.}(2022{\natexlab{b}})\citenamefont
  {Kim}, \citenamefont {Shi}, \citenamefont {Kato},\ and\ \citenamefont
  {Albert}}]{Kim_2022two}%
  \BibitemOpen
  \bibfield  {author} {\bibinfo {author} {\bibfnamefont {I.~H.}\ \bibnamefont
  {Kim}}, \bibinfo {author} {\bibfnamefont {B.}~\bibnamefont {Shi}}, \bibinfo
  {author} {\bibfnamefont {K.}~\bibnamefont {Kato}},\ and\ \bibinfo {author}
  {\bibfnamefont {V.~V.}\ \bibnamefont {Albert}},\ }\bibfield  {title}
  {\bibinfo {title} {Modular commutator in gapped quantum many-body systems},\
  }\href {https://doi.org/10.1103/PhysRevB.106.075147} {\bibfield  {journal}
  {\bibinfo  {journal} {Phys. Rev. B}\ }\textbf {\bibinfo {volume} {106}},\
  \bibinfo {pages} {075147} (\bibinfo {year} {2022}{\natexlab{b}})}\BibitemShut
  {NoStop}%
\bibitem [{\citenamefont {Shi}\ \emph {et~al.}(2020)\citenamefont {Shi},
  \citenamefont {Kato},\ and\ \citenamefont {Kim}}]{bootstrap}%
  \BibitemOpen
  \bibfield  {author} {\bibinfo {author} {\bibfnamefont {B.}~\bibnamefont
  {Shi}}, \bibinfo {author} {\bibfnamefont {K.}~\bibnamefont {Kato}},\ and\
  \bibinfo {author} {\bibfnamefont {I.~H.}\ \bibnamefont {Kim}},\ }\bibfield
  {title} {\bibinfo {title} {Fusion rules from entanglement},\ }\href
  {https://doi.org/https://doi.org/10.1016/j.aop.2020.168164} {\bibfield
  {journal} {\bibinfo  {journal} {Annals of Physics}\ }\textbf {\bibinfo
  {volume} {418}},\ \bibinfo {pages} {168164} (\bibinfo {year}
  {2020})}\BibitemShut {NoStop}%
\bibitem [{\citenamefont {Fan}\ \emph {et~al.}(2023)\citenamefont {Fan},
  \citenamefont {Zhang},\ and\ \citenamefont {Gu}}]{Fan_2023}%
  \BibitemOpen
  \bibfield  {author} {\bibinfo {author} {\bibfnamefont {R.}~\bibnamefont
  {Fan}}, \bibinfo {author} {\bibfnamefont {P.}~\bibnamefont {Zhang}},\ and\
  \bibinfo {author} {\bibfnamefont {Y.}~\bibnamefont {Gu}},\ }\bibfield
  {title} {\bibinfo {title} {Generalized real-space chern number formula and
  entanglement hamiltonian},\ }\bibfield  {journal} {\bibinfo  {journal}
  {SciPost Physics}\ }\textbf {\bibinfo {volume} {15}},\ \href
  {https://doi.org/10.21468/scipostphys.15.6.249}
  {10.21468/scipostphys.15.6.249} (\bibinfo {year} {2023})\BibitemShut
  {NoStop}%
\bibitem [{\citenamefont {Zou}\ \emph {et~al.}(2022)\citenamefont {Zou},
  \citenamefont {Shi}, \citenamefont {Sorce}, \citenamefont {Lim},\ and\
  \citenamefont {Kim}}]{Zou_mod_comm_CFT_2022}%
  \BibitemOpen
  \bibfield  {author} {\bibinfo {author} {\bibfnamefont {Y.}~\bibnamefont
  {Zou}}, \bibinfo {author} {\bibfnamefont {B.}~\bibnamefont {Shi}}, \bibinfo
  {author} {\bibfnamefont {J.}~\bibnamefont {Sorce}}, \bibinfo {author}
  {\bibfnamefont {I.~T.}\ \bibnamefont {Lim}},\ and\ \bibinfo {author}
  {\bibfnamefont {I.~H.}\ \bibnamefont {Kim}},\ }\bibfield  {title} {\bibinfo
  {title} {Modular commutators in conformal field theory},\ }\href
  {https://doi.org/10.1103/PhysRevLett.129.260402} {\bibfield  {journal}
  {\bibinfo  {journal} {Phys. Rev. Lett.}\ }\textbf {\bibinfo {volume} {129}},\
  \bibinfo {pages} {260402} (\bibinfo {year} {2022})}\BibitemShut {NoStop}%
\bibitem [{\citenamefont {Fan}(2022)}]{Fan_mod_comm_CFT_2022}%
  \BibitemOpen
  \bibfield  {author} {\bibinfo {author} {\bibfnamefont {R.}~\bibnamefont
  {Fan}},\ }\bibfield  {title} {\bibinfo {title} {From entanglement generated
  dynamics to the gravitational anomaly and chiral central charge},\ }\href
  {https://doi.org/10.1103/PhysRevLett.129.260403} {\bibfield  {journal}
  {\bibinfo  {journal} {Phys. Rev. Lett.}\ }\textbf {\bibinfo {volume} {129}},\
  \bibinfo {pages} {260403} (\bibinfo {year} {2022})}\BibitemShut {NoStop}%
\bibitem [{\citenamefont {Kitaev}\ and\ \citenamefont
  {Preskill}(2006)}]{kitaevpreskill2006}%
  \BibitemOpen
  \bibfield  {author} {\bibinfo {author} {\bibfnamefont {A.}~\bibnamefont
  {Kitaev}}\ and\ \bibinfo {author} {\bibfnamefont {J.}~\bibnamefont
  {Preskill}},\ }\bibfield  {title} {\bibinfo {title} {Topological entanglement
  entropy},\ }\href {https://doi.org/10.1103/PhysRevLett.96.110404} {\bibfield
  {journal} {\bibinfo  {journal} {Phys. Rev. Lett.}\ }\textbf {\bibinfo
  {volume} {96}},\ \bibinfo {pages} {110404} (\bibinfo {year}
  {2006})}\BibitemShut {NoStop}%
\bibitem [{\citenamefont {Levin}\ and\ \citenamefont
  {Wen}(2006)}]{levinwen2006}%
  \BibitemOpen
  \bibfield  {author} {\bibinfo {author} {\bibfnamefont {M.}~\bibnamefont
  {Levin}}\ and\ \bibinfo {author} {\bibfnamefont {X.-G.}\ \bibnamefont
  {Wen}},\ }\bibfield  {title} {\bibinfo {title} {Detecting topological order
  in a ground state wave function},\ }\href
  {https://doi.org/10.1103/PhysRevLett.96.110405} {\bibfield  {journal}
  {\bibinfo  {journal} {Phys. Rev. Lett.}\ }\textbf {\bibinfo {volume} {96}},\
  \bibinfo {pages} {110405} (\bibinfo {year} {2006})}\BibitemShut {NoStop}%
\bibitem [{\citenamefont {Bravyi}(2008)}]{BravyiUnpublished_2008}%
  \BibitemOpen
  \bibfield  {author} {\bibinfo {author} {\bibfnamefont {S.}~\bibnamefont
  {Bravyi}},\ }\bibfield  {title} {\bibinfo {title} {Unpublished},\ }\href@noop
  {} {\  (\bibinfo {year} {2008})}\BibitemShut {NoStop}%
\bibitem [{\citenamefont {Zou}\ and\ \citenamefont {Haah}(2016)}]{Zou_2016}%
  \BibitemOpen
  \bibfield  {author} {\bibinfo {author} {\bibfnamefont {L.}~\bibnamefont
  {Zou}}\ and\ \bibinfo {author} {\bibfnamefont {J.}~\bibnamefont {Haah}},\
  }\bibfield  {title} {\bibinfo {title} {Spurious long-range entanglement and
  replica correlation length},\ }\href
  {https://doi.org/10.1103/PhysRevB.94.075151} {\bibfield  {journal} {\bibinfo
  {journal} {Phys. Rev. B}\ }\textbf {\bibinfo {volume} {94}},\ \bibinfo
  {pages} {075151} (\bibinfo {year} {2016})}\BibitemShut {NoStop}%
\bibitem [{\citenamefont {Williamson}\ \emph {et~al.}(2019)\citenamefont
  {Williamson}, \citenamefont {Dua},\ and\ \citenamefont
  {Cheng}}]{Williamson_2019}%
  \BibitemOpen
  \bibfield  {author} {\bibinfo {author} {\bibfnamefont {D.~J.}\ \bibnamefont
  {Williamson}}, \bibinfo {author} {\bibfnamefont {A.}~\bibnamefont {Dua}},\
  and\ \bibinfo {author} {\bibfnamefont {M.}~\bibnamefont {Cheng}},\ }\bibfield
   {title} {\bibinfo {title} {Spurious topological entanglement entropy from
  subsystem symmetries},\ }\href
  {https://doi.org/10.1103/PhysRevLett.122.140506} {\bibfield  {journal}
  {\bibinfo  {journal} {Phys. Rev. Lett.}\ }\textbf {\bibinfo {volume} {122}},\
  \bibinfo {pages} {140506} (\bibinfo {year} {2019})}\BibitemShut {NoStop}%
\bibitem [{\citenamefont {Cano}\ \emph {et~al.}(2015)\citenamefont {Cano},
  \citenamefont {Hughes},\ and\ \citenamefont {Mulligan}}]{Cano_2015}%
  \BibitemOpen
  \bibfield  {author} {\bibinfo {author} {\bibfnamefont {J.}~\bibnamefont
  {Cano}}, \bibinfo {author} {\bibfnamefont {T.~L.}\ \bibnamefont {Hughes}},\
  and\ \bibinfo {author} {\bibfnamefont {M.}~\bibnamefont {Mulligan}},\
  }\bibfield  {title} {\bibinfo {title} {Interactions along an entanglement cut
  in $2+1\mathrm{D}$ abelian topological phases},\ }\href
  {https://doi.org/10.1103/PhysRevB.92.075104} {\bibfield  {journal} {\bibinfo
  {journal} {Phys. Rev. B}\ }\textbf {\bibinfo {volume} {92}},\ \bibinfo
  {pages} {075104} (\bibinfo {year} {2015})}\BibitemShut {NoStop}%
\bibitem [{\citenamefont {Fliss}\ \emph {et~al.}(2017)\citenamefont {Fliss},
  \citenamefont {Wen}, \citenamefont {Parrikar}, \citenamefont {Hsieh},
  \citenamefont {Han}, \citenamefont {Hughes},\ and\ \citenamefont
  {Leigh}}]{Fliss_2017}%
  \BibitemOpen
  \bibfield  {author} {\bibinfo {author} {\bibfnamefont {J.~R.}\ \bibnamefont
  {Fliss}}, \bibinfo {author} {\bibfnamefont {X.}~\bibnamefont {Wen}}, \bibinfo
  {author} {\bibfnamefont {O.}~\bibnamefont {Parrikar}}, \bibinfo {author}
  {\bibfnamefont {C.-T.}\ \bibnamefont {Hsieh}}, \bibinfo {author}
  {\bibfnamefont {B.}~\bibnamefont {Han}}, \bibinfo {author} {\bibfnamefont
  {T.~L.}\ \bibnamefont {Hughes}},\ and\ \bibinfo {author} {\bibfnamefont
  {R.~G.}\ \bibnamefont {Leigh}},\ }\bibfield  {title} {\bibinfo {title}
  {Interface contributions to topological entanglement in abelian chern-simons
  theory},\ }\bibfield  {journal} {\bibinfo  {journal} {Journal of High Energy
  Physics}\ }\textbf {\bibinfo {volume} {2017}},\ \href
  {https://doi.org/10.1007/jhep09(2017)056} {10.1007/jhep09(2017)056} (\bibinfo
  {year} {2017})\BibitemShut {NoStop}%
\bibitem [{\citenamefont {Santos}\ \emph {et~al.}(2018)\citenamefont {Santos},
  \citenamefont {Cano}, \citenamefont {Mulligan},\ and\ \citenamefont
  {Hughes}}]{Santos_2018}%
  \BibitemOpen
  \bibfield  {author} {\bibinfo {author} {\bibfnamefont {L.~H.}\ \bibnamefont
  {Santos}}, \bibinfo {author} {\bibfnamefont {J.}~\bibnamefont {Cano}},
  \bibinfo {author} {\bibfnamefont {M.}~\bibnamefont {Mulligan}},\ and\
  \bibinfo {author} {\bibfnamefont {T.~L.}\ \bibnamefont {Hughes}},\ }\bibfield
   {title} {\bibinfo {title} {Symmetry-protected topological interfaces and
  entanglement sequences},\ }\href {https://doi.org/10.1103/PhysRevB.98.075131}
  {\bibfield  {journal} {\bibinfo  {journal} {Phys. Rev. B}\ }\textbf {\bibinfo
  {volume} {98}},\ \bibinfo {pages} {075131} (\bibinfo {year}
  {2018})}\BibitemShut {NoStop}%
\bibitem [{\citenamefont {Stephen}\ \emph {et~al.}(2019)\citenamefont
  {Stephen}, \citenamefont {Dreyer}, \citenamefont {Iqbal},\ and\ \citenamefont
  {Schuch}}]{Stephen_2019}%
  \BibitemOpen
  \bibfield  {author} {\bibinfo {author} {\bibfnamefont {D.~T.}\ \bibnamefont
  {Stephen}}, \bibinfo {author} {\bibfnamefont {H.}~\bibnamefont {Dreyer}},
  \bibinfo {author} {\bibfnamefont {M.}~\bibnamefont {Iqbal}},\ and\ \bibinfo
  {author} {\bibfnamefont {N.}~\bibnamefont {Schuch}},\ }\bibfield  {title}
  {\bibinfo {title} {Detecting subsystem symmetry protected topological order
  via entanglement entropy},\ }\href
  {https://doi.org/10.1103/PhysRevB.100.115112} {\bibfield  {journal} {\bibinfo
   {journal} {Phys. Rev. B}\ }\textbf {\bibinfo {volume} {100}},\ \bibinfo
  {pages} {115112} (\bibinfo {year} {2019})}\BibitemShut {NoStop}%
\bibitem [{\citenamefont {Kato}\ and\ \citenamefont
  {Brand\~ao}(2020)}]{Kato_2020}%
  \BibitemOpen
  \bibfield  {author} {\bibinfo {author} {\bibfnamefont {K.}~\bibnamefont
  {Kato}}\ and\ \bibinfo {author} {\bibfnamefont {F.~G. S.~L.}\ \bibnamefont
  {Brand\~ao}},\ }\bibfield  {title} {\bibinfo {title} {Toy model of boundary
  states with spurious topological entanglement entropy},\ }\href
  {https://doi.org/10.1103/PhysRevResearch.2.032005} {\bibfield  {journal}
  {\bibinfo  {journal} {Phys. Rev. Res.}\ }\textbf {\bibinfo {volume} {2}},\
  \bibinfo {pages} {032005} (\bibinfo {year} {2020})}\BibitemShut {NoStop}%
\bibitem [{Note1()}]{Note1}%
  \BibitemOpen
  \bibinfo {note} {The operator $i \gamma P_{AB} P_{BC}$ is Hermitian since
  $P_{AB}, P_{BC}$ anticommute.}\BibitemShut {Stop}%
\bibitem [{\citenamefont {Briegel}\ and\ \citenamefont
  {Raussendorf}(2001)}]{Briegel_cluster_state_2001}%
  \BibitemOpen
  \bibfield  {author} {\bibinfo {author} {\bibfnamefont {H.~J.}\ \bibnamefont
  {Briegel}}\ and\ \bibinfo {author} {\bibfnamefont {R.}~\bibnamefont
  {Raussendorf}},\ }\bibfield  {title} {\bibinfo {title} {Persistent
  entanglement in arrays of interacting particles},\ }\href
  {https://doi.org/10.1103/PhysRevLett.86.910} {\bibfield  {journal} {\bibinfo
  {journal} {Phys. Rev. Lett.}\ }\textbf {\bibinfo {volume} {86}},\ \bibinfo
  {pages} {910} (\bibinfo {year} {2001})}\BibitemShut {NoStop}%
\bibitem [{Note2()}]{Note2}%
  \BibitemOpen
  \bibinfo {note} {To make $c_-$ well-defined, we can imagine embedding $|\psi
  \rangle $ in a 2D qubit lattice by putting every qubit in the lattice in a
  product state except for a 1D chain-like subset of $4N+1$ qubits which we put
  in the state $|\psi \rangle $.}\BibitemShut {Stop}%
\bibitem [{Note3()}]{Note3}%
  \BibitemOpen
  \bibinfo {note} {More precisely, since the blue gates are applied before the
  black gates, we need to remove them in the reverse order: first, we remove
  the \protect \emph {black} gates that do not intersect a boundary of $A$,
  $B$, $C$. Then we remove the blue gates that do not intersect one of these
  boundaries and do not touch a residual black gate.}\BibitemShut {Stop}%
\end{thebibliography}%
%
%%%%%%%%%%%%%%%%%%%%%%%%%%%%%%%%
%%%%%%%%%%%%%%%%%%%%%%%%%%%%%%%%

\begin{appendix}

%----------------------------------------------------------------------------------------------------------------
%----------------------------------------------------------------------------------------------------------------
%----------------------------------------------------------------------------------------------------------------
\section{Review of cluster states}
\label{sec:clusterstate}
In this Appendix, we review the definition of cluster states~\cite{Briegel_cluster_state_2001}. We also explain how cluster states can give rise to nonlocal modular Hamiltonians~\cite{BravyiUnpublished_2008, Zou_2016}.

\begin{figure}
\subfloat[\label{fig:clustergraph1}]{\def\svgwidth{4.0cm}%% Creator: Inkscape 1.3.2 (091e20e, 2023-11-25), www.inkscape.org
%% Accompanies image file 'clusterstabilizer_new.pdf' (pdf, eps, ps)
\begingroup%
  \makeatletter%
  \providecommand\color[2][]{%
    \errmessage{(Inkscape) Color is used for the text in Inkscape, but the package 'color.sty' is not loaded}%
    \renewcommand\color[2][]{}%
  }%
  \providecommand\transparent[1]{%
    \errmessage{(Inkscape) Transparency is used (non-zero) for the text in Inkscape, but the package 'transparent.sty' is not loaded}%
    \renewcommand\transparent[1]{}%
  }%
  \providecommand\rotatebox[2]{#2}%
  \newcommand*\fsize{\dimexpr\f@size pt\relax}%
  \newcommand*\lineheight[1]{\fontsize{\fsize}{#1\fsize}\selectfont}%
  \ifx\svgwidth\undefined%
    \setlength{\unitlength}{412.4172421bp}%
    \ifx\svgscale\undefined%
      \relax%
    \else%
      \setlength{\unitlength}{\unitlength * \real{\svgscale}}%
    \fi%
  \else%
    \setlength{\unitlength}{\svgwidth}%
  \fi%
  \global\let\svgwidth\undefined%
  \global\let\svgscale\undefined%
  \makeatother%
  \begin{picture}(1,0.78625382)%
    \lineheight{1}%
    \setlength\tabcolsep{0pt}%
    \put(0,0){\includegraphics[width=\unitlength,page=1]{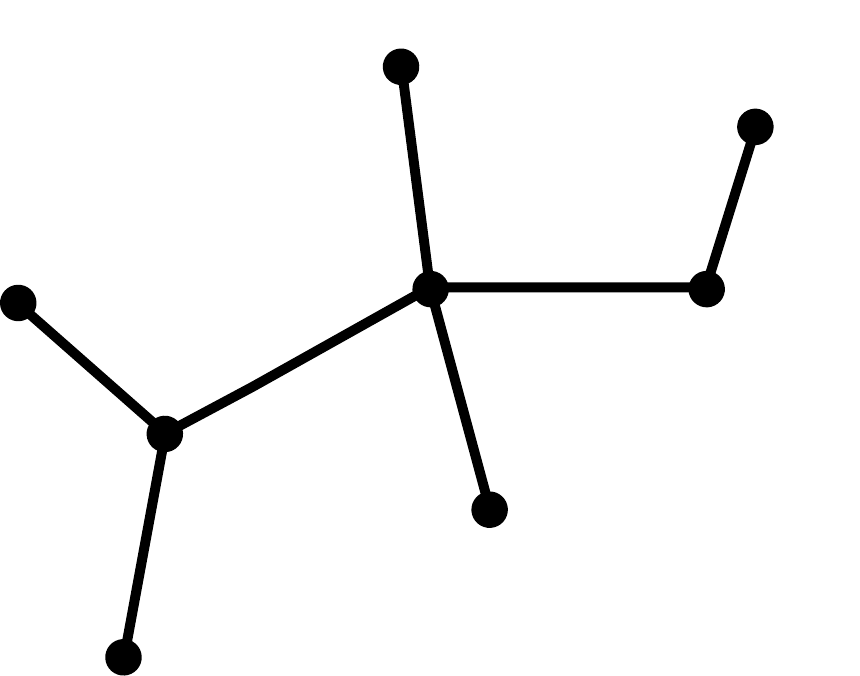}}%
    \put(0.52069687,0.46826357){\color[rgb]{0,0,0}\makebox(0,0)[lt]{\lineheight{1.25}\smash{\begin{tabular}[t]{l}$X$\end{tabular}}}}%
    \put(0.59637443,0.21600897){\color[rgb]{0,0,0}\makebox(0,0)[lt]{\lineheight{1.25}\smash{\begin{tabular}[t]{l}$Z$\end{tabular}}}}%
    \put(0.48839368,0.72855855){\color[rgb]{0,0,0}\makebox(0,0)[lt]{\lineheight{1.25}\smash{\begin{tabular}[t]{l}$Z$\end{tabular}}}}%
    \put(0.18795912,0.31756703){\color[rgb]{0,0,0}\makebox(0,0)[lt]{\lineheight{1.25}\smash{\begin{tabular}[t]{l}$Z$\end{tabular}}}}%
    \put(0.86100315,0.46826357){\color[rgb]{0,0,0}\makebox(0,0)[lt]{\lineheight{1.25}\smash{\begin{tabular}[t]{l}$Z$\end{tabular}}}}%
  \end{picture}%
\endgroup%
}
\hfill
\subfloat[\label{fig:clustergraph2}]{\def\svgwidth{4.0cm}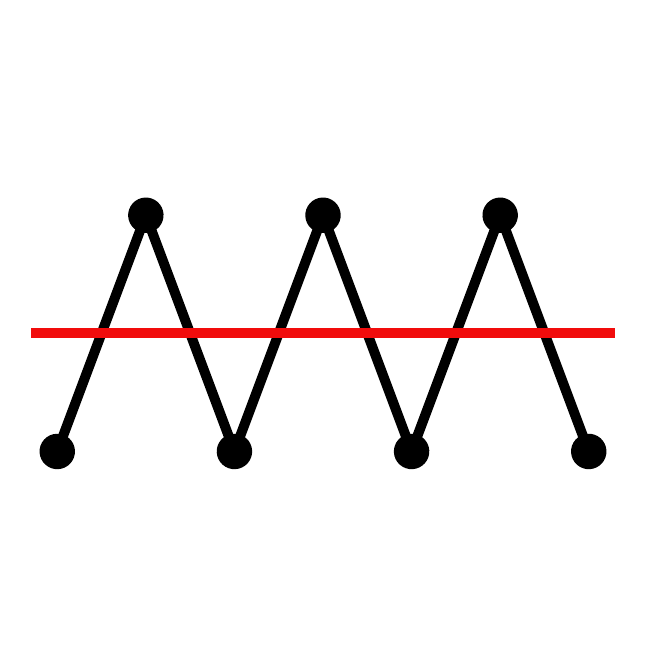}
%\subfloat[\label{fig:clustergraph1}]{\includegraphics[width=3.5cm]{clusterstabilizer.pdf}} 
%\hfill
%\subfloat[\label{fig:clustergraph2}]{\includegraphics[width=3.5cm]{1dcluster.pdf}} 
     \caption{(a) Cluster state operator $h_i$ for a general graph. (b) One-dimensional cluster state with open boundary conditions for $N=3$. The even-numbered $h_i$ multiply to give a term completely supported below the entanglement cut.}
\end{figure}

The cluster state is a many-body state that can be defined on any graph with vertices $V$ and edges $E$ where each vertex corresponds to a qubit. For each $i \in V$, we define local operators
\begin{align}
    \label{eq:hicluster}
    h_i = X_i\prod_{\<ij\>\in E}Z_j, 
\end{align}
where the product is taken over all qubits connected to $i$ by an edge (see Fig.~\ref{fig:clustergraph1}). Note that all the $h_i$ commute with one another so they can be simultaneously diagonalized. The cluster state is then defined as the state $|\psi_{cs}\>$ that satisfies $h_i|\psi_{cs}\>=|\psi_{cs}\>$ for every $i\in V$, i.e.~it is the ground state of the Hamiltonian
\begin{align}
\label{eq:clusterstate_H}
    H_{cs} = -\sum_{i\in V}h_i.
\end{align}

It is easy to see that $\psi_{cs}$ is unique by considering the projector onto the ground state space of $H_{cs}$
\begin{align}
\label{eq:clusterstate_rho}
    \rho_{cs}=\prod_{i\in V}\left(\frac{\id+h_i}{2}\right).
\end{align}
From the definition of the $h_i$, every term in $\rho_{cs}$ will be a nontrivial Pauli operator except for the term resulting from the product of the identities. Since this is the only term that contributes to the trace, we find that $\rho_{cs}$ has unit trace, and we conclude $\rho_{cs}=|\psi_{cs}\>\<\psi_{cs}|$. 

We now explain how cluster states can generate nonlocal modular Hamiltonians. As a specific example, consider the one-dimensional cluster state with open boundary conditions. Numbering the qubits from $0$ to $2N$, the density operator is given by
\begin{align}
\label{eq:1dclusterstate}
    \rho_{cs}^{1D} = \frac{1}{2^{2N+1}}(\id+X_0Z_1)&\prod_{i=1}^{2N-1}(\id+Z_{i-1}X_iZ_{i+1})\nonumber\\
    &\times (\id+Z_{2N-1}X_{2N}).
\end{align}
After expanding out the product, we see that $\rho_{cs}^{1D}$ will be a sum of Pauli string operators that nearly all have nontrivial support on both even and odd-numbered qubits. In fact the only term that has support only on the even-numbered qubits is that which results from pairing the two endpoint operators with the even-numbered bulk terms (see Fig.~\ref{fig:clustergraph2}),
\begin{align}
\label{eq:clusterchain}
    (X_0Z_1)(Z_1X_2Z_3)\cdots (Z_{2N-3}&X_{2N-2}Z_{2N-1})(Z_{2N-1}X_{2N})\nonumber\\
    &=X_0X_2\cdots X_{2N-2}X_{2N}\nonumber\\
    &\equiv\str{0}{2N}.
\end{align}
Therefore, tracing out the odd-numbered qubits leaves us with the reduced density operator
\begin{equation}
    \Tr_{\text{odd}}\rho_{cs}^{1D}=\frac{1}{2^{N+1}}(\id+\str{0}{2N}).
    %\Tr_{\text{odd}}\rho_{cs}^{1D}=\frac{1}{2^{N+1}}(\id+X_0X_2\cdots X_{2N}).
\end{equation}
To get the modular Hamiltonian, we need to take the logarithm of this expression. Strictly speaking, this logarithm is not well-defined since the density operator has vanishing eigenvalues. If we introduce a regulator by changing the coefficient of $\str{0}{2N}$ from $1 \to 1-\epsilon$ we obtain
\begin{align}
    -\ln \Tr_{\text{odd}}\rho_{cs}^{1D}= (N+1)\ln{2}\cdot\id&-\ln(2-\epsilon)\frac{\id+\str{0}{2N}}{2}\nonumber\\
    &-\ln(\epsilon)\frac{\id-\str{0}{2N}}{2}.
    %c \id +\frac{1}{2}\ln\left(\frac{2-\epsilon}{\epsilon}\right) X_0X_2\cdots X_{2N} 
    \label{eq:modhamcluster}
\end{align}
The modular Hamiltonian of the cluster state is therefore highly nonlocal for this particular choice of subregion, in the sense that it is a linear combination of the identity operator and the string $\str{0}{2N}$. (For readers who are troubled by the divergence of the last term as $\epsilon \rightarrow 0$, we note that this divergence drops out when computing most quantities of interest. For example, the last term does not contribute to the von Neumann entropy, since it has a vanishing expectation value in the state $\rho_{cs}$.)

More generally, long Pauli $X$ strings like $\str{0}{2N}$ arise generically in modular Hamiltonians of cluster state systems when the entanglement cut crosses a series of alternating lattice sites. They are at the heart of some of the simplest examples with spurious topological entanglement entropy~\cite{BravyiUnpublished_2008, Zou_2016}, and we similarly take advantage of this nonlocal structure to construct systems with spurious modular commutators. 
%----------------------------------------------------------------------------------------------------------------

\section{Entangling circuit for $|\psi\>$}
\label{app:1D_circuit}
In this Appendix, we describe an explicit depth-$2$ circuit that creates our 1D state $|\psi\>$ from a product state. Let $|\psi_+\>$ denote the simultaneous eigenstate of the $X_i$ operators with $X_i |\psi_+\> = |\psi_+\>$ for all $i$. We claim that $|\psi\> = \mathcal{C}|\psi_+\>$ where $\mathcal{C}$ is the following quantum circuit:
\begin{align}
\label{eq:circuit}
%|\psi\> &= C|\psi_+\>, \qquad \quad X_i|\psi_+\>=|\psi_+\>\ \  \forall i\nonumber\\
\mathcal{C} &= \left(\prod_{\substack{i=-2N,\\i\neq 0}}^{2N-1}\text{CZ}_{i,i+1}\right)\cdot V_{1,0}
%\biggl(\text{CNOT}_{1,0} \textbf{R}_0\biggr)
\end{align}
Here, $\text{CZ}_{i,j}$ is a controlled Z gate acting on $i,j$, while $V_{i,j}$ is a two-qubit gate defined by
\begin{align}
V_{i,j} = \text{CNOT}_{i,j} \mathbf{R}_j\
\label{eq:Vdef}
\end{align}
where $\text{CNOT}_{i,j}$ is the controlled NOT operator with $i$ as the control qubit, and $\mathbf{R}_j$ is any rotation operator acting on the $j$th qubit, with the property that $\mathbf{R}_j X_j \mathbf{R}_j^\dagger = \frac{1}{\sqrt{3}}(X_j + Y_j + Z_j)$.

We now demonstrate that the circuit $\mathcal{C}$ defined in (\ref{eq:circuit}) transforms $|\psi_+\>$ into $|\psi\>$. The initial product state has a density operator 
\begin{align*}
    \rho_+ = |\psi_+\>\<\psi_+| = \prod_{i=-2N}^{2N}\left(\frac{1+X_i}{2}\right).
\end{align*}
The rotation operator $\textbf{R}_j$ transforms $X_0 \rightarrow h_{0R}$, where the arrow denotes conjugation and
\begin{align*}
    h_{0R} = \frac{1}{\sqrt{3}}(X_0 + Z_0 + Y_0).
\end{align*}
Also, the $\text{CNOT}_{1,0}$ gate takes $X_{1}\rightarrow h_{1C}$ and $h_{0R}\rightarrow h_{0C}$, where
\begin{align*}
    h_{1C} &= X_0X_1\nonumber\\
    h_{0C} &= \frac{1}{\sqrt{3}}(X_0 + Z_0Z_1 + Y_0Z_1). 
\end{align*}
So after the first two-qubit unitary the state becomes 
\begin{align}
    \rho_+&\rightarrow V_{1,0} \rho_+ V_{1,0}^\dagger \nonumber\\
    %&= \frac{1}{2^{2(N+M)+1}}(1+X_{-2N})\cdots(1+h_{-1C})(1+h_{0C})\cdots(1+X_{2M})
    %&= (\frac{1+X_{-2N}}{2})\cdots(\frac{1+h_{-1C}}{2})(\frac{1+h_{0C}}{2})\cdots(\frac{1+X_{2M}}{2})
    &= \left(\prod_{\substack{i=-2N,\\i\neq 0,1}}^{2N-1}\frac{1+X_i}{2}\right)\left(\frac{1+h_{0C}}{2}\right)\left(\frac{1+h_{1C}}{2}\right)
\label{eq:rho+trans}
\end{align}
Finally, performing the controlled Z gate on every remaining pair of qubits has the effect of multiplying each $X_i$ operator by $Z_j$ for each of its neighbors and takes $Y_0\rightarrow Z_{-1} Y_0$. In other words, the $\text{CZ}$ gates take $X_i\rightarrow h_i$ and $h_{0C}\rightarrow h_0$ and $h_{1C}\rightarrow h_{1}$, where the $h_i$ are defined in (\ref{eq:hi})-(\ref{eq:h1}). Substituting these transformations into (\ref{eq:rho+trans}) we find that the circuit $\mathcal{C}$ indeed satisfies 
\begin{align}
    \mathcal{C}|\psi_+\>\<\psi_+|\mathcal{C} = |\psi\>\<\psi|.
\end{align}
%----------------------------------------------------------------------------------------------------------------

\section{Calculation of modular commutator for perturbed ground state}
\label{app:stability}
In this Appendix we compute the modular commutator $J(A,B,C)_{\tilde{\rho}}$ for the perturbed ground state $\tilde{\rho}$. The final result of this computation is Eq.~(\ref{eq:modcomperturbed}) from the main text. 

To begin, we reprint the key equations from the main text for the convenience of the reader. We recall that that the density operator $\tilde{\rho}$ is given by (\ref{eq:1dex_rho_pert}):
\begin{align}
    \tilde{\rho} = \prod_{i=-2N}^{2N}\left(\frac{\id+\tilde{h}_i}{2}\right).
\label{eq:1dex_rho_pert_app}
\end{align}
where $\tilde{h}_i = U(\theta) h_i U^\dagger(\theta)$. Also, we recall that $U(\theta)$ is given by (\ref{eq:Utheta}):
\begin{align}
    U(\theta) = \prod_{j=1}^{M}R_{-2j}(\theta) \prod_{k=1}^{M}R_{2k-1}(\theta)
\end{align}
where $R_i(\theta) = \exp(-i \theta Z_i Z_{i+1}/2)$. 

From the above expressions together with the definition of $h_i$ (\ref{eq:hi}-\ref{eq:h1}) we see that $\tilde{h}_i = h_i$ for $|i| \geq 2M+1$ since in this case $h_i$ commutes with $U(\theta)$. By the same reasoning, one can check that $\tilde{h}_0 = h_0$. Therefore, the only $\tilde{h}_i$ that differ from $h_i$ are those with $i\in [-2M,2M]\setminus \{0\}$. These $\tilde{h}_i$'s can be computed straightforwardly. For even $i \in [-2M, 2M] \setminus 0$, we find:
\begin{align}
\tilde{h}_i = \begin{cases} 
\cos{\theta}\ Z_{i-1}X_iZ_{i+1} + \sin{\theta}\ Z_{i-1}Y_i & i \leq -2 \\
\cos{\theta}\ Z_{i-1}X_{i}Z_{i+1} + \sin{\theta}\ Y_{i}Z_{i+1} & i \geq 2
\end{cases}
\label{eq:hieven}
\end{align}
Likewise, for odd $i \in [-2M, 2M] \setminus 0$, we find:
\begin{align}
\tilde{h}_i = \begin{cases} 
\cos{\theta}\ Z_{i-1}X_{i}Z_{i+1} + \sin{\theta}\ Y_{i}Z_{i+1} & i \leq -1 \\
\cos{\theta}\ Z_{-1}X_0X_1Z_2 + \sin{\theta}\ Z_{-1}X_0Y_1 & i = 1 \\
\cos{\theta}\ Z_{i-1}X_iZ_{i+1} + \sin{\theta}\ Z_{i-1}Y_i & i \geq 3
\end{cases}
\label{eq:hiodd}
\end{align}

We can now compute $\tilde{\rho}_{ABC}$ in the same way as in the unperturbed system: we expand out the product in (\ref{eq:1dex_rho_pert_app}) and trace out the odd numbered qubits. As in the unperturbed case, we can see that any term with an odd numbered $\tilde{h}_i$ will trace to zero, since every such $\tilde{h}_i$ contains an $X_i$ or $Y_i$ operator that cannot be cancelled by multiplication with other $h_j$'s. Therefore, we only need to consider products of even-numbered $\tilde{h}_i$. 

The main difference from the unperturbed case is that each even-number $\tilde{h}_i$ (\ref{eq:hieven}) is now a sum of two terms. Taking into account all the possible combinations of these terms that can contribute to the trace, we find that
\begin{align}
\label{eq:pabctilde}
    \tilde{\rho}_{ABC} = \frac{1}{2^{2N+1}}\biggl[\id&+\frac{1}{\sqrt{3}}(\cos^{M}\theta \str{-2N}{0} + \cos^{M}\theta Z_0\str{2}{2N} \nonumber\\
    &\qquad\qquad\qquad+ \cos^{2M}\theta \str{-2N}{-2}Y_0\str{2}{2N})\nonumber\\
    &+\sin{\theta}\sum_{j=0}^{M-1}\cos^j\theta \str{-2N}{-2(M-j)-2}Y_{-2(M-j)}\nonumber\\
    &+\sin{\theta}\sum_{k=0}^{M-1}\cos^k\theta Y_{2(M-k)}\str{2(M-k)+2}{2N}\biggr]
    %/2^{2N+1},
\end{align}
where we are using the notation defined in (\ref{eq:notation}). We now explain where each of the above Pauli strings come from. The first three Pauli strings in (\ref{eq:pabctilde}) are the same as in the unperturbed case (\ref{eq:1dpabc}) except for the additional factors of $\cos \theta$. These Pauli strings come from products of the $\cos \theta Z_{i-1} X_i  Z_{i+1}$ part of the $\tilde{h}_i$'s (\ref{eq:hieven}) matched with one of the three terms in $\tilde{h}_0 = h_0$. In other words, these terms originate in the same way as in the unperturbed case (\ref{eq:1dstrings}). The last two terms in (\ref{eq:pabctilde}) originate from 
taking a product of a string of $ \cos \theta Z_{i-1} X_i  Z_{i+1}$ terms terminated by one $ \sin \theta Y_i Z_{i \pm 1}$ term. This termination point can be at any even $i \in [-2M,2M] \setminus 0$, so this combination contributes many different Pauli strings to the trace. The first sum in (\ref{eq:pabctilde}) corresponds to the case where the termination point is at $i = -2(M-j)$ for $j=0,...,M-1$ while the the second sum in (\ref{eq:pabctilde}) corresponds to the case where the termination point is at $i = 2(M-k)$ where $k = 0,...,M-1$.

We now compute the reduced density operators $\tilde{\rho}_{AB}$ and $\tilde{\rho}_{BC}$. First note that the string operators that live in $AB$ -- the second and fifth terms in (\ref{eq:pabctilde}) -- mutually anticommute since the overlap between any two of them will contain one instance of $Y_{-2(M-j)}$. This means that the sum over all of these terms will square to a multiple of the identity. Therefore, we can write the reduced density operator $\tilde{\rho}_{AB}$ in the form 
\begin{align}
    \tilde{\rho}_{AB}= \frac{1}{2^{N+M+1}}(\id+\delta\widetilde{P}_{AB}),
\end{align}
where 
\begin{align*}
    \widetilde{P}_{AB} &= \frac{1}{\delta}\biggl( \frac{\cos^{M}\theta}{\sqrt{3}} \str{-2N}{0} \nonumber\\
    &\qquad\qquad + \sin{\theta}\sum_{j=0}^{M-1}\cos^j\theta \str{-2N}{-2(M-j)-2}Y_{-2(M-j)} \biggr)
\end{align*}
and where we choose $\delta$ so that $\widetilde{P}_{AB}^2 = \id$. In particular, $\delta$ is given by
\begin{align*}
    \delta^2 &= \left(\frac{\cos^{M}\theta}{\sqrt{3}}\right)^2 + \sin^2\theta\sum_{j=0}^{M-1}\cos^{2j}\theta\nonumber\\
    %&= \frac{\cos^{2M}\theta}{3} + \sin^2\theta\left(\frac{1-\cos^{2M}\theta}{1-\cos^2\theta}\right) 
    &= 1-\frac{2}{3}\cos^{2M}\theta.
\end{align*}
Since $\left(\widetilde{P}_{AB}\right)^2=1$, we can follow the same steps as in (\ref{eq:exampleK}) to find the modular Hamiltonian $\tilde{K}_{AB}$:
\begin{align}
    \tilde{K}_{AB} &= \tilde{c}_{AB}\id - \frac{1}{2}\ln\left(\frac{1+\delta}{1-\delta}\right) \widetilde{P}_{AB}
\end{align}
for some constant $\tilde{c}_{AB}$. We then repeat for region $BC$ to find 
\begin{align}
   \tilde{\rho}_{BC}= \frac{1}{2^{N + M + 1}}(\id+\delta\widetilde{P}_{BC}), 
\end{align}
where
\begin{align*}
    \widetilde{P}_{BC} &= \frac{1}{\delta}\biggl( \frac{\cos^{M}\theta}{\sqrt{3}} Z_0\str{2}{2N} \nonumber\\
    &\qquad\qquad + \sin{\theta}\sum_{k=0}^{M-1}\cos^k\theta Y_{2(M-k)}\str{2(M-k)+2}{2N} \biggr)
\end{align*}
Again, following the same steps as in (\ref{eq:exampleK}), we find
\begin{align}
    \tilde{K}_{BC} &= \tilde{c}_{BC}\id - \frac{1}{2}\ln\left(\frac{1+\delta}{1-\delta}\right) \widetilde{P}_{BC},
\end{align}
Taking the commutator $[\tilde{K}_{AB},\tilde{K}_{BC}]$, we see that the only terms that contribute are the original Pauli strings -- that is, the first terms in $\widetilde{P}_{AB}$ and $\widetilde{P}_{BC}$. In this way, we arrive at the modular commutator 
\begin{align}
    J(A,B,C)_{\tilde{\rho}} &= i\Tr(\tilde{\rho}_{ABC}[\tilde{K}_{AB},\tilde{K}_{BC}])\nonumber\\
    &=\frac{\cos^{4M}\theta}{6\sqrt{3}\delta^2}\ln^2\left(\frac{1+\delta}{1-\delta}\right).
\end{align}
This completes our derivation of Eq.~(\ref{eq:modcomperturbed}) from the main text.
%----------------------------------------------------------------------------------------------------------------
%----------------------------------------------------------------------------------------------------------------
%----------------------------------------------------------------------------------------------------------------

\end{appendix}

\end{document}